\documentclass[journal]{IEEEtran}
\usepackage{eurosym}
\usepackage{cite,algorithm,array,url}
\usepackage{graphicx,colortbl}
\usepackage{amsfonts,times,amsmath,amssymb,amsthm}

\theoremstyle{plain}
\newtheorem{theorem}{Theorem}
\newtheorem{lemma}[theorem]{Lemma}
\newtheorem{proposition}[theorem]{Proposition}

\theoremstyle{definition}

\theoremstyle{remark}
\newtheorem*{remark}{Remark}

\input{tcilatex}
\begin{document}

\title{An Efficient Implementation of the Generalized Labeled
Multi-Bernoulli Filter}
\author{Ba-Ngu~Vo, Ba-Tuong~Vo, Hung~Gia~Hoang\thanks{%
Acknowledgement: This work is supported by the Australian Research Council
under Discovery Project DP130104404.} \thanks{%
B.-N. Vo and B.-T. Vo are with the Department of Electrical and Computer
Engineering, Curtin University, Bentley, WA 6102, Australia (email:
\{ba-ngu.vo,ba-tuong.vo\}@curtin.edu.au). H.~G. Hoang is an independent
consultant (email: hghung@gmail.com)}}
\maketitle

\begin{abstract}
This paper proposes an efficient implementation of the generalized labeled
multi-Bernoulli (GLMB) filter by combining the prediction and update into a
single step. In contrast to an earlier implementation that involves separate
truncations in the prediction and update steps, the proposed implementation
requires only one truncation procedure for each iteration. Furthermore, we
propose an efficient algorithm for truncating the GLMB filtering density
based on Gibbs sampling. The resulting implementation has a linear
complexity in the number of measurements and quadratic in the number of
hypothesized objects.
\end{abstract}

\markboth{Preprint: IEEE Trans. Signal Processing,~Vol.~65,
No.~8, pp. 1975--1987, ~April~2017}{Vo et. al. : An Efficient Implementation of the Generalized Labeled
Multi-Bernoulli Filter}






\section{Introduction}

Multi-object tracking refers to the problem of jointly estimating the number
of objects and their trajectories from sensor data. Driven by aerospace
applications in the 1960's, today multi-object tracking lies at the heart of
a diverse range of application areas, see for example the texts \cite%
{BSF88,BP99, Mah07, mahler2014advances}. The most popular approaches to
multi-object tracking are the joint probabilistic data association filter
\cite{BSF88}, multiple hypothesis tracking \cite{BP99}, and recently, random
finite set (RFS) \cite{Mah07, mahler2014advances}.

The RFS approach has attracted significant attention as a general systematic
treatment of multi-object systems and provides the foundation for the
development of novel filters such as the Probability Hypothesis Density
(PHD) filter \cite{MahlerPHD}, Cardinalized PHD (CPHD) filter \cite%
{MahlerCPHD}, and multi-Bernoulli filters \cite{Mah07,VVC09,VVPS10}. While
these filters were not designed to estimate the trajectories of objects,
they have been successfully deployed in many applications including
radar/sonar \cite{TobiasLanterman05}, \cite{ClarkBell05}, computer vision
\cite{MTC_CSVT08,HVVS_PR12,HVV_TSP13}, cell biology \cite{Rez_TMI15},
autonomous vehicle \cite{MVA_TRO11, LHG_TSP11, LCS_SLAM_STSP13} automotive
safety \cite{Bat08,MRD13}, sensor scheduling \cite{RVC11,HV14sencon,
Gostar13, HVVM15}, and sensor network \cite%
{Zhang_TAC11,BCF_STSP13,UCJ_STSP13}.

The introduction of the generalized labeled multi-Bernoulli (GLMB) RFS in
\cite{VoGLMB13, VVP_GLMB13} has led to the development of the first
tractable and mathematically principled RFS-based multi-object tracker.
Recent extensions and applications \cite{ReuterLMB14, Gostar14, BVV15,
PapiKim15, Papi_etal15, Deusch15, Beard_etal16, Fantacci_etal16}, suggest
that the GLMB is a versatile model that offers good trade-offs between
tractability and fidelity. The GLMB filter exploits the conjugacy (with
respect to the standard measurement model) of the GLMB family to propagate
forward in time the (labeled) multi-object filtering density \cite{VoGLMB13}%
. Each iteration of this filter involves an update operation and a
prediction operation, both of which result in weighted sums of multi-object
exponentials with intractably large number of terms. The first
implementation of the GLMB filter truncates these sums by using the $K$%
-shortest path and ranked assignment algorithms, respectively, in the
prediction and update to determine the most significant components \cite%
{VVP_GLMB13}.

While the original two-staged implementation is intuitive and highly
parallelizable, it is structurally inefficient as two independent
truncations of the GLMB densities are required. Specifically, in the update,
truncation is performed by solving a ranked assignment problem for each
predicted GLMB component. Since truncation of the predicted GLMB sum is
performed separately from the update, a significant portion of the predicted
components would generate updated components with negligible weights. Thus,
computations are wasted in solving a large number of ranked assignment
problems with at best cubic complexity in the number of measurements.

In this paper, we present an efficient implementation of GLMB filter with
linear complexity in the number of measurements, i.e. at least two orders of
magnitude less than the original implementation in \cite{VVP_GLMB13}. In
particular, we derive a joint prediction and update that eliminates
inefficiencies in the truncation procedures of the original two-staged
implementation (this result has been presented at a conference in \cite%
{HVV15}). More importantly, we propose an efficient technique for truncating
the GLMB filtering density based on Gibbs sampling, which also offers an
efficient solution to the data association problem and more generally, the
ranked assignment problem. Further, we show that the proposed Gibbs sampler
has an exponential convergence rate. Naturally, in the joint prediction and
update, deterministic ranked assignment algorithms can also be applied to
truncate the GLMB filtering density. While both implementations are highly
parallelizable, the Gibbs sampler based solution has a linear complexity in
the number of measurements whereas deterministic solutions are cubic at best.

The paper is organized as follows. Background on labeled RFS and the GLMB
filter is provided in section \ref{sec:BG}. Section \ref{sec:fast_impl}
presents the joint prediction and update formulation and the Gibbs sampler
based implementation of the GLMB filter. Numerical results are presented in
Section \ref{sec:sim} and concluding remarks are given in Section \ref%
{sec:sum}.

\section{Background}

\label{sec:BG} This section summarizes the GLMB filter and its
implementation. The reader is referred to the original works \cite{VoGLMB13,
VVP_GLMB13} for detailed expositions.

Throughout this article, we denote a generalization of the Kroneker delta
that takes arbitrary arguments such as sets, vectors, integers etc., by
\begin{equation*}
\delta _{Y}[X]\triangleq \left\{
\begin{array}{l}
1,\text{ if }X=Y \\
0,\text{ otherwise}%
\end{array}%
\right. .
\end{equation*}%
The list of variables $X_{m},X_{m+1},...,X_{n}$ is abbreviated as $X_{m:n}$.
For a given set $S$, $1_{S}(\cdot )$ denotes the indicator function of $S$,
and $\mathcal{F}(S)$ denotes the class of finite subsets of $S$. For a
finite set $X$, its cardinality (or number of elements) is denoted by $|X|$,
in addition we use the multi-object exponential notation $f^{X}$ for the
product $\tprod_{x\in X}f(x)$, with $f^{\emptyset }=1$. The inner product $%
\int f(x)g(x)dx$ is denoted by $\left\langle f,g\right\rangle $.

\subsection{Labeled RFS\label{subsec:LabeledRFS}}

From a Bayesian estimation viewpoint the multi-object state is naturally
represented as a finite set, and subsequently modeled as an RFS or a
simple-finite point process \cite{MahlerPHD}. In this paper we use Mahler's
Finite Set Statistics (FISST) notion of integration/density (which is
consistent with measure theoretic integration/density \cite{VSD05}) to
characterize RFSs. Treatments of RFS in the context of multi-object
filtering can be found in \cite{Mah07, mahler2014advances}.

Consider a state space $\mathbb{X}$, and a discrete space $\mathbb{L}$, let $%
\mathcal{L}:\mathbb{X}\mathcal{\times }\mathbb{L}\rightarrow \mathbb{L}$ be
the projection defined by $\mathcal{L}((x,\ell ))=\ell $. Then $\mathcal{L}(%
\mathbf{x})$ is called the label of the point $\mathbf{x}\in \mathbb{X}%
\mathcal{\times }\mathbb{L}$, and a finite subset $\mathbf{X}$ of $\mathbb{X}%
\mathcal{\times }\mathbb{L}$ is said to have \emph{distinct labels} if and
only if $\mathbf{X}$ and its labels $\mathcal{L}(\mathbf{X})=\{\mathcal{L}(%
\mathbf{x}):\mathbf{x}\in \mathbf{X}\}$ have the same cardinality. The \emph{%
distinct label indicator} is defined by
\begin{equation*}
\Delta (\mathbf{X})\triangleq \delta _{|\mathbf{X}|}[|\mathcal{L(}\mathbf{X}%
)|]
\end{equation*}

A \emph{labeled RFS} is a marked simple point process with state space $%
\mathbb{X}$ and (discrete) mark space $\mathbb{L}$ such that each
realization has distinct labels \cite{VoGLMB13, VVP_GLMB13}. The distinct
labels provide the means to identify trajectories or tracks of individual
objects since a trajectory is a time-sequence of states with the same label.

A \textit{GLMB} is a labeled RFS with state space $\mathbb{X}$ and
(discrete) label space $\mathbb{L}$ distributed according to \cite{VoGLMB13,
VVP_GLMB13}
\begin{equation}
\mathbf{\pi }(\mathbf{X})=\Delta (\mathbf{X})\sum_{\xi \in \Xi }w^{(\xi )}(%
\mathcal{L}(\mathbf{X}))\left[ p^{(\xi )}\right] ^{\mathbf{X}}
\label{eq:glmb}
\end{equation}%
where $\Xi $ is a given discrete set, each $p^{(\xi )}(\cdot ,\ell )$ is a
probability density on $\mathbb{X}$, and each $w^{(\xi )}(L)$ is
non-negative with $\sum_{\xi \in \Xi }\sum_{L\in \mathcal{F}\!\left( \mathbb{%
L}\right) }w^{\left( \xi \right) }\!\left( L\right) =1$. Each term in the
mixture (\ref{eq:glmb}) consists of: a weight $w^{\left( \xi \right)
}\!\left( \mathcal{L}\!\left( \mathbf{X}\right) \right) $ that only depends
on the labels of the multi-object state $\mathbf{X}$; and a multi-object
exponential $\left[ p^{(\xi )}\right] ^{\mathbf{X}}$ that depends on the
entire multi-object state. A salient feature of the GLMB family is its
closure under the multi-object Bayes recursion for the standard multi-object
transition kernel and likelihood function \cite{VoGLMB13}.

Throughout the paper, single-object states are represented by lowercase
letters (e.g. $x$, $\mathbf{x}$) while multi-object states are represented
by uppercase letters (e.g. $X$, $\mathbf{X}$), symbols for labeled states
and their distributions are bolded (e.g. $\mathbf{x}$, $\mathbf{X}$, $%
\mathbf{\pi }$) to distinguish them from unlabeled ones, spaces are
represented by blackboard bold (e.g. $\mathbb{X}$, $\mathbb{Z}$, $\mathbb{L}$%
).

\subsection{Multi-object system model}

Using the convention from \cite{VoGLMB13}, an object is labeled by an
ordered pair $\ell =(k,i)$, where $k$ is the \textit{time of birth}, and $%
i\in \mathbb{N}$ is a unique index to distinguish objects born at the same
time. Thus, the label space for objects born at time $k$ is $\mathbb{B}_{k}%
\mathbb{\triangleq }\{k\}\times \mathbb{N}$, and an object born at time $k$
has state $\mathbf{x}\in \mathbb{X}\mathcal{\times }\mathbb{B}_{k}$.
Moreover, the label space $\mathbb{L}_{k}$ for objects at time $k$
(including those born prior to $k$) is given by $\mathbb{L}_{k}=\mathbb{L}%
_{k-1}\cup \mathbb{B}_{k}$ (note that $\mathbb{L}_{k-1}$ and $\mathbb{B}_{k}$
are disjoint). A multi-object state $\mathbf{X}$, at time $k$, is a finite
subset of $\mathbb{X}\mathcal{\times }\mathbb{L}_{k}$.

For compactness we omit the subscript $k$ for the current time, the next
time is indicated by the subscripts `$+$'.

Given the multi-object state $\mathbf{X}$ (at time $k$), each state $(x,\ell
)\in \mathbf{X}$ either survives with probability $P_{S}(x,\ell )$ and
evolves to a new state $(x_{+},\ell _{+})$ (at time $k+1$) with probability
density $f_{+}(x_{+}|x,\ell )\delta _{\ell }[\ell _{+}]$ or dies with
probability $1-P_{S}(x,\ell )$. The set $\mathbf{Y}$ of new targets born at
time $k+1$ is distributed according to the labeled multi-Bernoulli (LMB)
\begin{equation}
\Delta (\mathbf{Y})\left[ 1_{\mathbb{B}_{\,+}}\,r_{B,+}\right] ^{\mathcal{L(}%
\mathbf{Y})}\left[ 1-r_{B,+}\right] ^{\mathbb{B}_{+}-\mathcal{L(}\mathbf{Y}%
)}p_{B,+}^{\mathbf{Y}},  \label{eq:LMB_birth}
\end{equation}%
where $r_{B,+}(\ell _{+})$ is probability that a new object with label $\ell
_{+}$ is born, and $p_{B,+}(\cdot ,\ell _{+})$\ is the distribution of its
kinematic state \cite{VoGLMB13}. The multi-object state $\mathbf{X}_{+}$ (at
time $k+1$) is the superposition of surviving objects and new born objects.
It is assumed that, conditional on $\mathbf{X}$, \ objects move, appear and
die independently of each other. The expression for the multi-object
transition density can be found in \cite{VoGLMB13, VVP_GLMB13}.

For a given multi-object state $\mathbf{X}$ with distinct labels, each state
$(x,\ell )\in \mathbf{X}$ is either detected with probability $P_{D}(x,\ell
) $ and generates an observation $z$ with likelihood $g(z|x,\ell )$ or
missed with probability $1-P_{D}(x,\ell )$. The \emph{multi-object
observation} at time $k$, $Z=\{z_{1:|Z|}\}$, is the superposition of the
observations from detected objects and Poisson clutter with intensity $%
\kappa $. Assuming that, conditional on $\mathbf{X}$ (with distinct labels),
detections are independent of each other and of clutter, the multi-object
likelihood is given by \cite{VoGLMB13, VVP_GLMB13}
\begin{equation}
g(Z|\mathbf{X})\propto \sum_{\theta \in \Theta }1_{\Theta (\mathcal{L(}%
\mathbf{X}))}(\theta )\prod\limits_{(x,\ell )\in \mathbf{X}}\psi
_{Z_{_{\!}}}^{(\theta (\ell ))}(x,\ell )  \label{eq:RFSmeaslikelihood0}
\end{equation}%
where: $\Theta $ is the set of \emph{positive 1-1} maps $\theta :\mathbb{L}%
\rightarrow \{0$:$|Z|\}$, i.e. maps such that \emph{no two distinct labels
are mapped to the same positive value}; $\Theta (I)\subseteq \Theta $
denotes the set of positive 1-1 maps with domain $I$; and
\begin{equation}
\psi _{\{z_{1:|Z|}\}}^{(j)}(x,\ell )=\left\{ \!%
\begin{array}{ll}
\!\frac{P_{D}(x,\ell )g(_{\!}z_{j\!}|_{_{\!}}x,\ell )}{\kappa (z_{j})}, &
\!\!\text{if }j\in \left\{ 1,...,|Z|\right\} \\
\!1-P_{D}(x,\ell ), & \!\!\text{if }j=0%
\end{array}%
\right. \!\!.  \label{eq:PropConj5}
\end{equation}%
The map $\theta $ specifies which objects generated which measurements, i.e.
object $\ell $ generates measurement $z_{\theta (\ell )}\in Z$, with
undetected objects assigned to $0$. The positive 1-1 property means that $%
\theta $ is 1-1 on $\{\ell :\theta (\ell )>0\}$, the set of labels that are
assigned positive values, and ensures that any measurement in $Z$ is
assigned to at most one object.

\subsection{GLMB filter}

The GLMB filter propagates the multi-object filtering density forward in
time, analytically, under the multi-object transition and measurement
models. In implementation, the GLMB filtering density is expressed in an
alternative form\footnote{%
obtained by substituting $w^{(\xi )}(J)=\sum_{I\in \mathcal{F}\!(\mathbb{L}%
)}w^{(\xi )}(I)\delta _{I}(J)$ into (\ref{eq:glmb}).}, known as $\delta $%
-GLMB
\begin{equation}
\mathbf{\pi }(\mathbf{X})=\Delta (\mathbf{X})\sum_{\xi \in \Xi ,I\in
\mathcal{F}(\mathbb{L})}\omega ^{(I,\xi )}\delta _{I}[\mathcal{L(}\mathbf{X}%
)]\left[ p^{(\xi )}\right] ^{\mathbf{X}},  \label{eq:delta-glmb}
\end{equation}%
where $\omega ^{(I,\xi )}=w^{(\xi )}(I)$. Each $\xi \in \Xi $ represents a
history of association maps $\xi =(\theta _{1:k})$ while each $I\in \mathcal{%
F}(\mathbb{L})$ represents a set of object labels. Collectively, the weight $%
\omega ^{(I,\xi )}$ and function $p^{(\xi )}$ is called component $(I,\xi )$
of the $\delta $-GLMB.

Given the $\delta $-GLMB filtering density (\ref{eq:delta-glmb}) at time $k$%
, the $\delta $-GLMB prediction density to time $k+1$ is given by \cite%
{VoGLMB13}\footnote{%
Eq. (33) with the sum over $J$ $\in \mathcal{F}(I)$ replaced by the sum over
$\mathcal{F}(\mathbb{L})$ weighted by $1_{\!\mathcal{F}(I)\!}(_{\!}J_{\!})$,
and equating the sum over $I$ with $\bar{\omega}_{+}^{(J,L_{+\!},_{\!}\xi )}$%
.}%
\begin{equation}
\mathbf{\bar{\pi}}_{_{\!}+}(_{_{\!}}\mathbf{X})=\Delta _{{\!}}(_{_{\!}}%
\mathbf{X}_{_{\!}})\!\sum_{\xi ,J_{\!},L_{+}}\bar{\omega}_{+}^{(\xi
,J,L_{+\!})}\delta _{_{\!}J\cup L_{+\!}}[\mathcal{L}(\mathbf{X})]\!\left[
\bar{p}_{+}^{(\xi )}\right] ^{\mathbf{X}}  \label{eq:GLMB_pred0}
\end{equation}%
where $\xi \in \Xi $, $J\in \mathcal{F}(\mathbb{L})$, $L_{+}\in \mathcal{F}(%
\mathbb{B}_{+})$, and \allowdisplaybreaks%
\begin{eqnarray}
\!\!\!\bar{\omega}_{+}^{(\xi ,J,L_{+\!})}\!\!\! &=&\!\!\!1_{\mathcal{F}(%
\mathbb{B}_{{+}})}(L_{+})\text{ }r_{B\!,+}^{L_{\!+\!}}\left[ 1-r_{B,+}\right]
^{\mathbb{B}_{+}-L_{\!+\!}}  \notag \\
\! &\!\times &\!\!\!\!\sum_{I\in \mathcal{F}(\mathbb{L})}\!\!1_{\mathcal{F}%
(I)\!}(_{\!}J)\!\left[ _{\!}\bar{P}_{S}^{(\xi )\!}\right] ^{\!J}\!\left[
_{\!}1\!-\!\bar{P}_{S\!}^{_{\!}(_{\!}\xi )\!}\right] ^{\!I-J\!}\omega
^{(\!I,\xi )}  \label{eq:GLMB_pred1} \\
\!\!\!\bar{P}_{_{\!}S\!}^{(\xi )}(\ell )\!\!\! &=&\!\!\!\left\langle p^{(\xi
)}(\cdot ,\ell ),P_{S}(\cdot ,\ell )\right\rangle  \label{eq:GLMB_pred2} \\
\!\!\!\bar{p}_{+}^{(\xi )}(x_{+},\ell _{+})\!\!\! &=&\!\!\!1_{\mathbb{L}%
_{\!}}(\ell _{+})\frac{\!\left\langle P_{S}(\cdot ,\ell
_{+})f_{_{\!}+\!}(x_{+}|\cdot ,\ell _{+}),p^{(\xi )}(\cdot ,\ell
_{+})\right\rangle }{\bar{P}_{S}^{(\xi )}(\ell _{+})}  \notag \\
\! &+&\!\!\!1_{\mathbb{B}_{+\!}}(\ell _{+})p_{B,+_{_{\!}}}(x_{+},\ell _{+}).
\label{eq:GLMB_pred3}
\end{eqnarray}%
Moreover, the $\delta $-GLMB filtering density at time $k+1$ is \cite%
{VoGLMB13}\footnote{%
Eq. (13) with the sum over ${\Theta }_{\!+}(J\cup L_{\!+\!})$ replaced by
the sum over ${\Theta }_{\!+}$ weighted by $1_{_{\!}{\Theta }_{\!+}(J\cup
L_{\!+\!})}(_{\!}\theta _{\!+\!})$.}%
\begin{equation}
\mathbf{\pi }_{\!Z_{_{\!}+}\!}(\mathbf{X})\propto \Delta _{\!}(_{\!}\mathbf{X%
}_{\!})\!\!\!\sum\limits_{\xi ,J_{\!},L_{\!+\!},\theta _{\!+\!}}\!\!\!\omega
_{Z_{_{\!}+}}^{(_{\!}\xi ,J,L_{\!+\!},\theta _{\!+\!})}\delta _{J\cup
L_{\!+\!}}[\mathcal{L}(_{\!}\mathbf{X}_{\!})]\!\left[ p_{Z_{_{\!}+_{%
\!}}}^{_{\!}(_{\!}\xi ,\theta _{\!+\!})}{}_{\!}\right] ^{\mathbf{X}}\!\!
\label{eq:GLMB_upd0}
\end{equation}%
where $\theta _{+}\in \Theta _{+}$, and \allowdisplaybreaks%
\begin{eqnarray}
\!\!\!\!\!\!\!\!\omega _{Z_{_{\!}+}}^{(\xi ,J,L_{+\!},\theta _{+\!})}\!\!\!
&=&\!\!\!1_{{\Theta }_{+\!}(J\cup L_{\!+\!})}(\theta _{+})\!\!\left[ \bar{%
\psi}_{Z_{+}}^{(\xi ,\theta _{\!+\!})\!}\right] ^{\!J\cup L_{+}}\!\!\bar{%
\omega}_{+}^{(\xi ,J,L_{+})}  \label{eq:GLMB_upd1} \\
\!\!\!\!\!\!\!\!\bar{\psi}_{Z_{+}}^{(\xi ,\theta _{_{\!}+_{\!}})}(\ell
_{_{\!}+_{\!}})\!\!\! &=&\!\!\!\left\langle \bar{p}_{+}^{(\xi )}(\cdot ,\ell
_{+}),\psi _{_{\!}Z_{_{\!}+}}^{(\theta _{+}(\ell _{+}))}(\cdot ,\ell
_{+})\right\rangle  \label{eq:GLMB_upd2} \\
\!\!\!\!\!\!\!\!p_{Z_{+}}^{(\xi ,\theta
_{_{\!}+_{\!}})_{\!}}(x_{_{\!}+},\ell _{_{\!}+_{\!}})\!\!\! &=&\!\!\!\frac{%
\bar{p}_{+}^{(\xi )}(x_{+},\ell _{+})\psi _{Z_{+}}^{(\theta _{+}(\ell
_{+}))}(x_{+},\ell _{+})}{\bar{\psi}_{Z_{+}}^{(\xi ,\theta _{+})}(\ell _{+})}
\label{eq:GLMB_upd3}
\end{eqnarray}

\begin{remark}
To be concise, $\bar{\psi}_{Z_{+}}^{(\xi ,\theta _{+})}(\ell _{+})$ should
be understood as $\bar{\psi}_{Z_{_{\!}+}}^{(\xi ,\theta _{+}(\ell
_{+}))}(\ell _{_{\!}+})$ since (\ref{eq:GLMB_upd2}) does not require
knowledge of the entire map $\theta _{+}$, but only its value at $\ell _{+}$%
. Similarly, $p_{Z_{_{\!}+}}^{(\xi ,\theta _{+})}(x_{+},\ell _{+})$ should
be understood as $p_{Z_{+}}^{(\xi ,\theta _{+}(\ell _{+}))}(x_{+},\ell _{+})$%
.
\end{remark}

The first implementation of the GLMB filter recursively computes the
predicted and update densities at each time step \cite{VVP_GLMB13}. Since
the number of components in the $\delta $-GLMB prediction and filtering
densities grows exponentially with time, these densities are truncated by
retaining components with largest weights to minimize the $L_{1}$ truncation
error \cite{VVP_GLMB13}.

From the $\delta $-GLMB weight prediction (\ref{eq:GLMB_pred0})-(\ref%
{eq:GLMB_pred1}), note that each component $(I,\xi )$ generates a new set of
\textquotedblleft children\textquotedblright\ $(I,\xi ,J,L_{+})$, $%
J\subseteq I$, $L_{+}\subseteq \mathbb{B}_{+}$, with weights proportional to
$[\bar{P}_{S}^{(\xi )}]^{J}[~1~\!~-~\!~\bar{P}_{S\!}^{(_{\!}\xi
)}~]^{I\!-\!J}$ $r_{B\!,+\!}^{L_{+}}\left[ 1-r_{B\!,+}\right] ^{\mathbb{B}%
_{+}-L_{+}}$. Truncating the contribution of component $(I,\xi )$ to the
predicted density without exhaustively computing all the children's weights
is performed by solving two separate $K$-shortest path problems. The first
finds a prescribed number of surviving label sets $J\subseteq I$ in
non-increasing order of $[\bar{P}_{S}^{(\xi )}]^{J}[1-\bar{P}_{S\!}^{(\xi
)}]^{I\!-\!J}$, while the second finds a prescribed number of new label sets
$L_{+}\subseteq \mathbb{B}_{+}$, in non-increasing order of $%
r_{B,+\!}^{L_{+}}\left[ 1-r_{B\!,+\!}\right] ^{\mathbb{B}_{+}-L_{\!+\!}}$.
This strategy ensures that components with births, which usually have very
low weights, are retained.

From the $\delta $-GLMB weight update (\ref{eq:GLMB_upd0})-(\ref%
{eq:GLMB_upd1}), note that each prediction component $(\xi ,J,L_{+})$
generates a new set of components $(\xi ,J,L_{+},\theta _{+})$, $\theta
_{+}\in \Theta _{+}(J\cup L_{+})$, with weights proportional to $[\bar{\psi}%
_{Z_{+}}^{(\xi ,\theta _{+})}]^{J\cup L_{+}}$. Truncating the contribution
of component $(\xi ,J,L_{+})$ to the updated density without exhaustively
computing all the components is performed by solving the ranked assignment
problem to find a prescribed number of association maps $\theta _{+}$ in
non-increasing order of $[\bar{\psi}_{Z_{_{\!}+}}^{(\xi ,\theta
_{+})}]^{J\cup L_{+}}$.

Separating the truncation of the prediction from the update only exploits
\textit{a priori} knowledge (e.g. survival and birth probabilities),
consequently, a significant portion of these predicted components would
generate updated components with negligible weights. Hence, computations are
wasted in solving a large number of ranked assignment problems, which have,
at best, cubic complexity in the number of measurements.

\section{Efficient implementation of the GLMB filter}

\label{sec:fast_impl}

To avoid propagating prediction components that would generate weak updated
components, this section proposes a new implementation of the GLMB filter
with a joint prediction and update. This joint strategy only requires one
truncation procedure per iteration, while preserving the filtering
performance as well as parallelizability. Further, we detail a $\delta $%
-GLMB truncation approach based on Gibbs sampling that drastically reduces
the complexity.

The $\delta $-GLMB joint prediction and update is presented in subsection~%
\ref{subsec:new_scheme}, followed by a formulation of the $\delta $-GLMB
truncation problem in subsection \ref{subsec:truncform}. The ranked
assignment and Gibbs sampling solutions to the truncation problem are given
in subsections \ref{subsec:RankedAss} and \ref{subsec:fast_assginment},
followed by implementations details in subsection \ref{subsec:implementation}%
.

\subsection{Joint prediction and update}

\label{subsec:new_scheme}

The following Proposition establishes a direct recursion between the
components of two GLMB filtering densities at consecutive times (see the
Appendix for proof).

\begin{proposition}
\label{Prop_joint} Given the $\delta $-GLMB filtering density (\ref%
{eq:delta-glmb}) at time $k$, the $\delta $-GLMB filtering density at time $%
k+1$ is given by \allowdisplaybreaks%
\begin{equation}
\mathbf{\pi }_{\!Z_{_{\!}+}\!}(\mathbf{X})\!\propto \Delta _{\!}(_{\!}%
\mathbf{X}_{\!})\!\!\!\!\sum\limits_{I\!,\xi ,I_{\!+},\theta
_{\!+\!}}\!\!\!\!\omega ^{(I,\xi )}\omega _{Z_{_{\!}+}}^{(_{\!}I_{\!},\xi
,I_{\!+\!},\theta _{\!+\!})}\delta _{_{\!}I_{+\!}}[\mathcal{L}(_{\!}\mathbf{X%
}_{\!})]\!\left[ p_{Z_{_{\!}+_{\!}}}^{(_{\!}\xi ,\theta _{\!+\!})}{}_{\!}%
\right] ^{\!\mathbf{X}}\!\!\!  \label{eq:GLMB_joint0}
\end{equation}%
where $I\in \mathcal{F}(\mathbb{L})$, $\xi \in \Xi $, $I_{+}\in \mathcal{F}(%
\mathbb{L}_{+}),\theta _{+}\in \Theta _{+}$, and\allowdisplaybreaks%
\begin{eqnarray}
\!\!\!\!\!\!\omega _{Z_{_{\!}+}}^{(_{\!}I_{\!},\xi ,I_{\!+\!},\theta
_{\!+\!})}\!\!\! &=&\!\!\!1_{{\Theta }_{\!+\!}(I_{+})}(\theta _{\!+\!})\!%
\left[ 1-\bar{P}_{S}^{(\xi )}\right] ^{\!I\!-I_{\!+}}\!\left[ \bar{P}%
_{S\!}^{(\xi )}\right] ^{\!I\cap I_{+\!}}  \notag \\
\!\!\!\! &\times &\!\!\!\left[ 1-r_{B\!,+}\right] ^{\mathbb{B}%
_{\!+\!}-I_{\!+}\!}r_{B\!,+}^{\mathbb{B}_{\!{+}}\cap I_{+\!}}\!\left[ \bar{%
\psi}_{_{\!}Z_{_{\!}+}}^{(_{\!}\xi ,\theta _{_{\!}+\!})}\right] ^{I_{+}}
\label{eq:GLMB_joint1} \\
\!\!\!\!\!\!\bar{P}_{S\!}^{(\xi )}(\ell )\!\!\! &=&\!\!\!\left\langle
p^{(\xi )\!}(\cdot ,\ell ),P_{S}(\cdot ,\ell )\right\rangle
\label{eq:GLMB_joint2} \\
\!\!\!\!\!\!\bar{\psi}_{_{\!}Z_{_{\!}+}}^{(\xi ,\theta _{+\!})}(\ell
_{_{\!}+\!})\!\!\! &=&\!\!\!\left\langle \bar{p}_{+}^{(\xi )}(\cdot ,\ell
_{_{\!}+}),\psi _{_{\!}Z_{_{\!}+}\!}^{(\theta _{_{\!}+}(\ell
_{_{\!}+}))}(\cdot ,\ell _{_{\!}+})\right\rangle  \label{eq:GLMB_joint3} \\
\!\!\!\!\!\!\bar{p}_{+}^{(\xi )_{\!}}(x_{_{\!}+},\ell _{_{\!}+\!})\!\!\!
&=&\!\!\!1_{\mathbb{L}_{\!}}(\ell _{_{\!}+\!})\frac{\!\left\langle
P_{S}(\cdot ,\ell _{_{\!}+\!})f_{_{\!}+\!}(x_{_{\!}+}|\cdot ,\ell
_{_{\!}+\!}),p^{(\xi )}(\cdot ,\ell _{_{\!}+\!})\right\rangle }{\bar{P}%
_{S}^{(\xi )}(\ell _{_{\!}+})}  \notag \\
\!\!\!\! &+&\!\!\!1_{\mathbb{B}_{+}}\!(\ell
_{_{\!}+_{\!}})p_{B,+_{_{\!}}}(x_{_{\!}+},\ell _{_{\!}+})
\label{eq:GLMB_joint4} \\
\!\!\!\!\!\!p_{Z_{_{\!}+}}^{(\xi _{\!},\theta _{\!+\!})\!}(x_{_{\!}+},\ell
_{_{\!}+\!})\!\!\! &=&\!\!\!\frac{\bar{p}_{+}^{(\xi )}(x_{_{\!}+},\ell
_{_{\!}+})\psi _{Z_{+}}^{(\theta _{_{\!}+}(\ell _{_{\!}+}))}(x_{_{\!}+},\ell
_{_{\!}+})}{\bar{\psi}_{Z_{+}}^{(\xi ,\theta _{_{\!}+})}(\ell _{_{\!}+})}.
\label{eq:GLMB_joint5}
\end{eqnarray}
\end{proposition}

The summation (\ref{eq:GLMB_joint0}) can be interpreted as an enumeration of
all possible combinations of births, deaths and survivals together with
associations of new measurements to hypothesized labels. Observe that (\ref%
{eq:GLMB_joint0}) does indeed take on the $\delta $-GLMB form when rewritten
as a sum over $I_{+},\xi ,\theta _{+}$\ with weights
\begin{equation}
\omega _{Z_{+}}^{(I_{+},\xi ,\theta _{+})}\propto \sum\limits_{I}\omega
^{(I,\xi )}\omega _{Z_{+}}^{(I,\xi ,I_{+},\theta _{+\!})}.
\label{eq:GLMB_joint6}
\end{equation}%
Hence at the next iteration we only propagate forward the components $%
(I_{+},\xi ,\theta _{+})$ with weights $\omega _{Z_{_{\!}+}}^{(I_{+},\xi
,\theta _{+})}$.

The number of components in the $\delta $-GLMB filtering density grows
exponentially with time, and needs to be truncated at every time step,
ideally, by retaining those with largest weights since this minimizes the $%
L_{1}$ approximation error \cite{VVP_GLMB13}. Note from the $\delta $-GLMB
recursion (\ref{eq:GLMB_joint0})-(\ref{eq:GLMB_joint1}) that each component $%
(I,\xi )$ generates a set of \textquotedblleft children\textquotedblright\ $%
(I,\xi ,I_{+},\theta _{+})$, $(I_{+},\theta _{+})\in \mathcal{F}(\mathbb{L}%
_{+})\times \Theta _{+}(I_{+})$ with weights proportional to $\omega
_{Z_{_{\!}+}}^{(I,\xi ,I_{+},\theta _{+})}$. Truncating the contribution of
component $(I,\xi )$ to the $\delta $-GLMB filtering density, at time $k+1$,
amounts to selecting its \textquotedblleft children\textquotedblright\ $%
(I,\xi ,I_{+},\theta _{+})$ with significant weights. \vspace{-0.2cm}

\subsection{GLMB truncation formulation\label{subsec:truncform}}

In this and the next two subsections, we consider a fixed component $(\xi ,$
$I)$ of the $\delta $-GLMB filtering density at time $k$, and a fixed
measurement set $Z_{+}$ at time $k+1$. Specifically, we enumerate $%
Z_{+}=\{z_{1:M}\}$, $I=\{\ell _{1:R}\}$, and in addition $\mathbb{B}_{{+}%
\!}=\{\ell _{R+1:P}\}$. The goal is to find a set of pairs $(I_{+},\theta
_{+})\in \mathcal{F}(\mathbb{L}_{+})\times \Theta _{+}(I_{+})$ with
significant $\omega _{Z_{_{\!}+}}^{(I,\xi ,I_{+},\theta _{+})}$.

\begin{figure*}[t]
\centering\includegraphics[scale=.80]{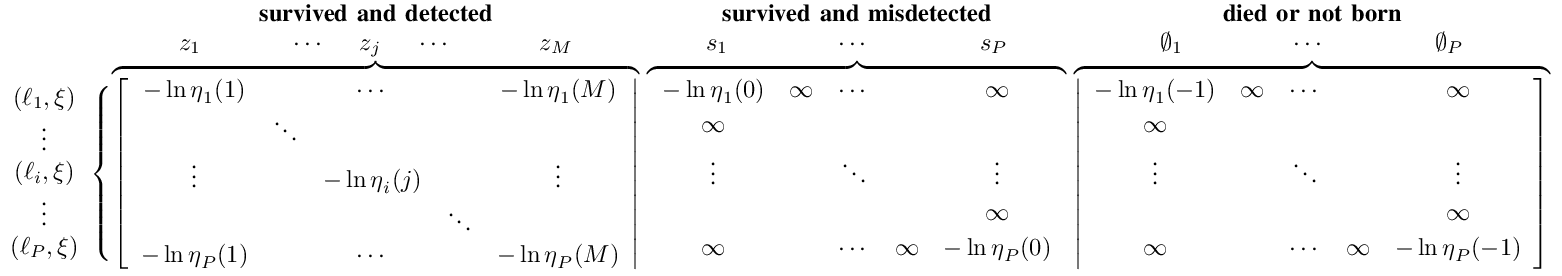}
\caption{The cost matrix $C$ for the joint prediction and update optimal
assignment problem of component $(\protect\xi ,$ $I)$. The 3 partitions
correspond to survived and detected objects, survived and misdetected
objects, and objects that died or not born. The assignment matrix $S$ has
the same structure, but with 1's and 0's as entries.}
\label{fig:costmat}
\end{figure*}

For each pair $(I_{+},\theta _{+})\in \mathcal{F}(\mathbb{L}_{+})\times
\Theta _{+}(I_{+})$, we define a $P$-tuple $\gamma =(\gamma _{1:P})\in \{-1$:%
$M\}^{P}$ by
\begin{equation*}
\gamma _{i}=\left\{
\begin{array}{ll}
\theta _{+}(\ell _{i}), & \text{if }\ell _{i}\in I_{+} \\
-1, & \text{otherwise}%
\end{array}%
\right.
\end{equation*}%
Note that $\gamma $ inherits, from $\theta _{+}$, the \emph{positive 1-1}
property, i.e., there are no distinct $i$, $i^{\prime }\in \{1$:$P\}$ with $%
\gamma _{i}\!=\!\gamma _{i^{\prime }}>0$. The set of all positive 1-1
elements of $\{-1$:$M\}^{P}$ is denoted by ${\Gamma }$. From $\gamma \in {%
\Gamma }$, we can recover $I_{+}$ and $\theta _{+}:I_{+}\rightarrow \{0$:$%
M\} $, respectively, by%
\begin{equation}
I_{+}=\{\ell _{i}\in I\cup \mathbb{B}_{\!{+}\!}:\gamma _{i}\geq 0\}\text{
and }\theta _{+}(\ell _{i})=\gamma _{i}.  \label{eq:convert}
\end{equation}%
Thus, $1_{{\Gamma }}(\gamma )=1_{_{\!}{\Theta }_{+}(I_{+})}(\theta _{+})$,
and there is a 1-1 correspondence between the spaces $\Theta _{+}(I_{+})$
and ${\Gamma }$.

Assuming that for all $i\in \{1$:$P\}$, $\bar{P}_{S}^{(\xi )\!}(\ell
_{i})\in (0,1)$ and $\bar{P}_{_{\!}D}^{(\xi )\!}(\ell _{i})\triangleq
\left\langle \bar{p}_{+}^{(\xi )\!}(\cdot ,\ell _{i}),P_{_{\!}D}(\cdot ,\ell
_{i})\right\rangle \in (0,1)$, we define
\begin{equation}
\eta _{i}(j)\!=\!%
\begin{cases}
1-\bar{P}_{S}^{(\xi )\!}(\ell _{i}), & \!1\leq i\leq R,\text{ }j\!<0\!, \\
\bar{P}_{S\!}^{(\xi )\!}(\ell _{i})\bar{\psi}_{Z_{+_{\!}}}^{(\xi ,j)\!}(\ell
_{i\!}), & \!1\leq i\leq R,\text{ }j\!\geq 0, \\
1-r_{B\!,+}(\ell _{i}), & \!R\!+\!1\leq i\leq P,\text{ }j\!<0, \\
r_{B\!,+}(\ell _{i})\bar{\psi}_{Z_{+}}^{(\xi ,j)}(\ell _{i}), &
\!R\!+\!1\leq i\leq P,\text{ }j\!\geq 0.%
\end{cases}
\label{eq:eta}
\end{equation}%
where $\bar{\psi}_{Z_{+}}^{(\xi ,j)}(\ell _{i})\!=\!\left\langle \bar{p}%
_{+}^{(\xi )}(\cdot ,\ell _{i}),\psi _{_{\!}Z_{+}}^{(j)}(\cdot ,\ell
_{i})\right\rangle $, and $j\in \{-1$:$M\}$ is the index of the measurement
assigned to label $\ell _{i}$, with $j=0$ indicating that $\ell _{i}$ is
misdetected, and $j=-1$ indicating that $\ell _{i}$ no longer exists. It is
implicit that $\eta _{i}(j)$ depends on the given $(\xi ,$ $I)$ and $Z_{+}$,
which have been omitted for compactness. The assumptions on the expected
survival and detection probabilities, $\bar{P}_{S}^{(\xi )\!}(\ell _{i})$
and $\bar{P}_{D}^{(\xi )\!}(\ell _{i})$, eliminates trivial and ideal
sensing scenarios, as well as ensuring $\eta _{i}(j)>0$.

Note from (\ref{eq:convert}) that since $\theta _{+}(\ell _{i})$ = $\gamma
_{i}$, we have $\bar{\psi}_{Z_{+}}^{(\xi ,\gamma _{i})}(\ell _{i})$ = $\bar{%
\psi}_{Z_{+}}^{(\xi ,\theta _{+}(\ell _{i}))}(\ell _{i})$ = $\bar{\psi}%
_{Z_{+}}^{(\xi ,\theta _{+})}(\ell _{i})$, (see (\ref{eq:GLMB_upd2}) and the
subsequent remark), hence it follows from (\ref{eq:eta}) that
\begin{eqnarray*}
\prod\limits_{n=1}^{R}\eta _{n}(\gamma _{n})\!\! &=&\!\!\left[ 1-\bar{P}%
_{S}^{(\xi )}\right] ^{I-I_{+}}\left[ \bar{P}_{S}^{(\xi )}\bar{\psi}%
_{Z_{+}}^{(\xi ,\theta _{+})}\right] ^{I\cap I_{+}}\!, \\
\prod\limits_{n=R+1}^{P}\!\eta _{n}(\gamma _{n})\!\! &=&\!\!\left[
1-r_{B\!,+}\right] ^{\mathbb{B}_{+}-I_{+}}\left[ r_{B\!,+}\bar{\psi}%
_{Z_{+}}^{(\xi ,\theta _{+})}\right] ^{\mathbb{B}_{{+}}\cap I_{+}}\!.
\end{eqnarray*}%
Moreover, using (\ref{eq:GLMB_joint1}), we have
\begin{equation}
\omega _{Z_{+}}^{(I,\xi ,I_{+},\theta _{+})}=1_{{\Gamma }}(\gamma
)\prod\limits_{i=1}^{P}\eta _{i}(\gamma _{i}).  \label{eq:eta_prod}
\end{equation}%
Consequently, \emph{finding a set of} $(I_{+},\theta _{+})\in \mathcal{F}(%
\mathbb{L}_{+})\times \Theta _{+}(I_{+})$ \emph{with significant} $\omega
_{Z_{_{\!}+}}^{(I,\xi ,I_{+},\theta _{+})}$ \emph{is equivalent to finding a
set of positive 1-1 vectors} $\gamma $ \emph{with significant} $%
\tprod\nolimits_{i=1\!}^{P}\eta _{i}(\gamma _{i})$.

\subsection{Ranked Assignment\label{subsec:RankedAss}}

Similar to the GLMB update implementation in \cite{VVP_GLMB13}, for a given
component $(I,\xi )$, the $T$ best positive 1-1 vectors $\gamma $ in
non-increasing order of $\prod\nolimits_{i=1}^{P}\eta _{i}(\gamma _{i})$,
can be obtained, without exhaustive enumeration, by solving the following
ranked assignment problem.

Each $\gamma \in {\Gamma }$ can be represented by a $P\!\times \!(M+2P)$
\emph{assignment matrix} $S$ consisting of $0$ or $1$ entries with every row
summing to $1$, and every column summing to either $1$ or $0$. Note that $S$
can be partitioned into 3 sub-matrices similar to the matrix shown in Fig. %
\ref{fig:costmat}. For $(i,j)$ $\in $ $\{1$:$P\}\!\times \!\{1$:$M\}$, $%
S_{i,j}=1$ when $\gamma _{i}$ $=$ $j$ (i.e. $\ell _{i}$ generates the $j$th
measurement). For $(i,j)$ $\in $ $\{1$:$P\}\!\times \!\{M+1$:$M+P\}$, $%
S_{i,j}=1$ when $\gamma _{i}$ $=$ $0$ and $j=M+i$ (i.e. $\ell _{i}$ not
detected). For $(i,j)$ $\in $ $\{1$:$P\}\!\times \!\{M\!+\!P\!+\!1$:$%
M\!+\!2P\}$, $S_{i,j}=1$ when $\gamma _{i}$ $=$ $-1$ and $j=M\!+\!P\!+\!i$
(i.e. $\ell _{i}$ does not exist). More concisely,
\begin{equation*}
S_{i,j}=1_{\{1:M\}}(j)\delta _{\gamma _{i\!}}[j]+\delta _{\!M+i}[j]\delta
_{\gamma _{i\!}}[0]+\delta _{\!M+P+i}[j]\delta _{\gamma _{i\!}}[-1].
\end{equation*}

The $P\!\times \!(M+2P)$ \emph{cost matrix }$C$ of this optimal assignment
problem (implicitly depends on $(\xi ,$ $I)$ and $Z_{+}$), is given by%
\begin{equation}
C_{i,j}=%
\begin{cases}
-\ln \eta _{i}(j) & j\in \{1:M\} \\
-\ln \eta _{i}(0) & j=M+i \\
-\ln \eta _{i}(-1) & j=M+P+i \\
\infty & \text{otherwise}%
\end{cases}
\label{eq:Ass_Cost_Matrix1}
\end{equation}%
(see also Fig. \ref{fig:costmat}). The cost of an assignment matrix $S$ is
\begin{equation*}
\text{tr}(S^{T}C)=\sum_{i=1}^{P}\sum_{j=1}^{M+2P}C_{i,j}S_{i,j},
\end{equation*}%
and is related to the weight of the corresponding positive 1-1 vector $%
\gamma $ by $\exp \!\left( -\text{tr}(S^{T}C)\right)
=\tprod\nolimits_{i=1\!}^{P}\eta _{i}(\gamma _{i})$.

Note that the cost matrix (\ref{eq:Ass_Cost_Matrix1}) is an extension of the
cost matrix of the GLMB update implementation in \cite{VVP_GLMB13} to
integrate birth, death, survival, detection, misdetection and clutter. A
GLMB filter implementation using Murty's algorithm to solve the above ranked
assignment problem has been reported in \cite{HVV15}. The same strategy of
solving ranked assignment problems with joint prediction and update was
proposed for unlabeled multi-object filtering in \cite{Corea15}. Although
this approach does not produce tracks like the GLMB filter, it is still
useful in applications such as mapping \cite{MVA_TRO11, LHG_TSP11} where the
individual trajectories of the landmarks are not required.

Solving the ranked assignment problem with cost matrix $C$, for the $T$ best
positive 1-1 vectors can be accomplished by Murty's algorithm \cite{Murty68}
with a complexity of $\mathcal{O}\left( T(M+2P)^{4}\right) $. More efficient
algorithms \cite{Milleretal97}, \cite{Pedersenetal08} can reduce the
complexity to $\mathcal{O}\left( T(M+2P)^{3}\right) $. The main contribution
of this article is a much cheaper and simpler algorithm for generating
positive 1-1 vectors with high weights. 

\subsection{Gibbs Sampling}

\label{subsec:fast_assginment}

The main drawback in using existing ranked assignment algorithms are the
high computational cost of generating a sequence of positive 1-1 vectors
ordered according to their weights, whilst such ordering is not needed in
the $\delta $-GLMB approximation. In this subsection, we propose a more
efficient alternative by using Markov Chain Monte Carlo (MCMC) to simulate
an unordered set of significant positive 1-1 vectors. In particular, we
exploit the Gibbs sampler to break down a complex high-dimensional problem
into simple, low-dimensional problems to achieve greater efficiency.

The key idea in the stochastic simulation approach is to consider $\gamma $
as a realization of a random variable distributed according to a probability
distribution $\pi $ on $\{-1$:$M\}^{P}$. Candidate positive 1-1 vectors are
then generated by independently sampling from $\pi $. Asymptotically, the
proportion of samples with probabilities above a given threshold is equal to
the probability mass of points with probabilities above that threshold. To
ensure that mostly high-weight positive 1-1 vectors are sampled, $\pi $ is
constructed so that only positive 1-1 vectors have positive probabilities,
and those with high weights are more likely to be chosen than those with low
weights. An obvious choice of $\pi $ is one that assigns each positive 1-1
vector a probability proportional to its weight, i.e.
\begin{equation}
\pi (\gamma )\propto 1_{{\Gamma }}(\gamma )\prod\limits_{i=1\!}^{P}\eta
_{i}(\gamma _{i})  \label{eq:thetajoint_dis}
\end{equation}%
where ${\Gamma }$ is the set of positive 1-1 vectors in $\{-1$:$M\}^{P}$.

Sampling directly from the distribution (\ref{eq:thetajoint_dis}) is very
difficult. MCMC is a widely used technique for sampling from a complex
distribution by constructing a suitable Markov chain. Indeed, MCMC
simulation has been applied to compute posterior distributions of data
association variables in multi-object tracking problems \cite{Oh}. However,
depending on the proposal, it could take some time for a new sample to be
accepted. Designing a proposal to have high acceptance probability is still
an open area of research. Furthermore, the actual distribution of the
samples from a Markov chain depends on the starting value, even if
asymptotically the samples are distributed according to the stationary
distribution. Usually, an MCMC simulation is divided into two parts: the
pre-convergence samples, known as burn-ins, are discarded; and the
post-convergence samples are used for inference \cite{ComparisonMCMC05}. The
key technical problem is that there are no bounds on the burn-in time nor
reliable techniques for determining when convergence has occurred, see e.g.
\cite{ComparisonMCMC05} and references therein.

The Gibbs sampler is a computationally efficient special case of the
Metropolis-Hasting MCMC algorithm, in which proposed samples are always
accepted \cite{Geman_Gibbs84,Cassella_Gibbs92}. Further, in GLMB filtering,
we are not interested in the distribution of the positive 1-1 vectors.
Regardless of their distribution, all distinct positive 1-1 vectors will
reduce the $L_{1}$ GLMB approximation error. Thus, unlike MCMC posterior
inference, there are no problems with burn-ins.

\vspace{7pt}

\hrule

\vspace{5pt}

\textbf{Algorithm 1. Gibbs }

\begin{itemize}
\item \textsf{{\footnotesize {input: }}}$\gamma ^{(1)},T,\eta =[\eta
_{i}(j)] $

\item \textsf{{\footnotesize {output: }}}$\gamma ^{(1)},...,\gamma ^{(T)}$
\end{itemize}

\vspace{2pt}

\hrule

\vspace{4pt}

$P:=\mathsf{size}(\eta ,1);\quad M:=\mathsf{size}(\eta ,2)-2;$

\textsf{{\footnotesize {for}}\ }$t=2:T$

\quad \textsf{{\footnotesize {for }}}$n=1:P$

\quad \quad $\gamma _{n}^{(t)}\sim \pi _{n}(\cdot |\gamma
_{1:n-1}^{(t)},\gamma _{n+1:P}^{(t-1)})$;

\quad \textsf{{\footnotesize {end}}}

\quad $\gamma ^{(t)}:=[\gamma _{1}^{(t)},...,\gamma _{P}^{(t)}];$

\textsf{{\footnotesize {end}}}

\vspace{5pt}

\hrule

\vspace{7pt}

Formally, the Gibbs sampler (see Algorithm 1) is a Markov chain $\{\gamma
^{(t)}\}_{t=1}^{\infty }$ with transition kernel \cite%
{Geman_Gibbs84,Cassella_Gibbs92}
\begin{equation*}
\pi (\gamma ^{\prime }|\gamma )=\dprod\limits_{i=1}^{P}\pi _{n}(\gamma
_{n}^{\prime }|\gamma _{1:n-1}^{\prime },\gamma _{n+1:P}),
\end{equation*}%
where $\pi _{n}(\gamma _{n}^{\prime }|\gamma _{1:n-1}^{\prime },\gamma
_{n+1:P})\propto \pi (\gamma _{1:n}^{\prime },\gamma _{n+1:P})$. In other
words, given $\gamma =(\gamma _{1:P})$, the components $\gamma _{1}^{\prime
},...,\gamma _{P}^{\prime }$ of the state at the next iterate of the chain,
are distributed according to the sequence of conditionals%
\begin{align*}
\pi _{1}(\gamma _{1}^{\prime }|\gamma _{_{\!}2:P})& \propto \pi (\gamma
_{1}^{\prime },\gamma _{_{\!}2:P}) \\
& \text{ \ }\vdots \\
\pi _{n}(\gamma _{n}^{\prime }|\gamma _{1:n-1}^{\prime },\gamma _{n+1:P})&
\propto \pi (\gamma _{1:n}^{\prime },\gamma _{n+1:P}) \\
& \text{ \ }\vdots \\
\pi _{P}(\gamma _{P}^{\prime }|\gamma _{1:P-1}^{\prime })& \propto \pi
(\gamma _{1:P}^{\prime }).
\end{align*}%
Although the Gibbs sampler is computationally efficient with an acceptance
probability of 1, it requires the conditionals $\pi _{n}(\cdot |\cdot )$, $%
n\in \{1$:$P\}$, to be easily computed and sampled from.

In the following we establish closed form expressions for the conditionals
that can be computed/sampled at low cost.

\begin{lemma}
\label{lemma} Let $\bar{n}=\{1$:$P\}-\{n\}$, $\gamma _{\bar{n}\!}=(\gamma
_{_{\!}1:n_{\!}-_{\!}1\!},\gamma _{_{\!}n_{\!}+_{\!}1:P})$, and ${\Gamma }(%
\bar{n})$ be the set of all positive 1-1 $\gamma _{\bar{n}}$ (i.e. $\gamma _{%
\bar{n}}$ for which there are no distinct $i,j\in \bar{n}$ with $\gamma
_{i}\!=\!\gamma _{j}\!>\!0$). Then, for any $\gamma \in \{-1$:$M\}^{P}$, $1_{%
{\Gamma }}(\gamma )$ can be factorized as:%
\begin{equation}
1_{{\Gamma }}(\gamma )=1_{{\Gamma }(\bar{n})}(\gamma _{\bar{n}%
\!})\dprod\limits_{i\in \bar{n}}(1-1_{\{1:M\}}(\gamma _{n})\delta _{\gamma
_{n}}[\gamma _{i}]).  \label{eq:LemmaGibbs}
\end{equation}
\end{lemma}

The proof is given in the Appendix.

\begin{proposition}
\label{marg_cond} For each $n\in \{1$:$P\}$,
\begin{equation}
\pi _{n}(\gamma _{n}|\gamma _{\bar{n}})\propto \eta _{n}(\gamma
_{n})\dprod\limits_{i\in \bar{n}}(1-1_{\{1:M\}}(\gamma _{n})\delta _{\gamma
_{n}}[\gamma _{i}]).  \label{eq:marg_cond1}
\end{equation}
\end{proposition}

\textbf{Proof}: We are interested in highlighting the functional dependence
of $\pi _{n}(\gamma _{n}|\gamma _{\bar{n}})$ on $\gamma _{n}$, while its
dependence on all other variables is aggregated into the normalizing
constant:
\begin{eqnarray*}
\pi _{n}(\gamma _{n}|\gamma _{\bar{n}}) &\triangleq &\frac{\pi (\gamma )}{%
\pi (\gamma _{\bar{n}})}\propto \pi (\gamma )\propto 1_{{\Gamma }}(\gamma
)\prod\limits_{j=1}^{P}\eta _{j}(\gamma _{j}) \\
&=&\eta _{n}(\gamma _{n})1_{{\Gamma }}(\gamma )\prod\limits_{j\in \bar{n}%
}\eta _{j}(\gamma _{j}).
\end{eqnarray*}%
Factorizing $1_{{\Gamma }}(\gamma )$ using Lemma \ref{lemma}, gives
\begin{eqnarray*}
&&\!\!\!\!\!\!\!\!\!\!\!\!\!\!\!\!\!\!\!\!\pi _{n}(\gamma _{n}|\gamma _{\bar{%
n}}) \\
\!\! &\propto &\!\!\eta _{_{\!}n_{\!}}(\gamma
_{n})_{\!\!}\dprod\limits_{i\in \bar{n}}(1\!-\!1_{\{1:M\}\!}(\gamma
_{n})\delta _{\!\gamma _{n}}[\gamma _{i}])1_{{\Gamma }\!(\bar{n})\!}(\gamma
_{\!\bar{n}})\!\prod\limits_{j\in \bar{n}}\eta _{j}(\gamma _{j}) \\
\!\! &\propto &\!\!\eta _{_{\!}n_{\!}}(\gamma
_{n})_{\!\!}\dprod\limits_{i\in \bar{n}}(1\!-\!1_{\{1:M\}\!}(\gamma
_{n})\delta _{\!\gamma _{n}}[\gamma _{i}]).\text{ \ \ \ \ \ \ \ \ \ \ \ \ \
\ \ \ \ \ \ }\square
\end{eqnarray*}

For a non-positive $j$, $1_{\{1:M\}\!}(j)=0$, and Proposition~\ref{marg_cond}
implies $\pi _{n}(j|\gamma _{\bar{n}})$ $\propto $ $\eta _{n}(j)$. On the
other hand, given any $j$ $\in $ $\{1:M\}$, Proposition~\ref{marg_cond}
implies that $\pi _{n}(j|\gamma _{\bar{n}})$ $\propto $ $\eta _{n}(j)$,
unless $j\in \{\gamma _{1:n-1},\gamma _{n+1:P}\}$, i.e. there is an $i\in
\bar{n}$ with $\gamma _{i}=j$, in which case $\pi _{n}(j|\gamma _{\bar{n}%
})=0 $. Consequently, for $j$ $\in $ $\{1:M\}$
\begin{equation*}
\pi _{n}(j|\gamma _{\bar{n}})\propto \eta _{n}(j)(1-1_{\{\gamma
_{1:n-1},\gamma _{n+1:P}\}}(j)).
\end{equation*}%
Hence, sampling from the conditionals is simple and inexpensive as
illustrated in Algorithm 1a, which has an $\mathcal{O}(PM)$ complexity since
sampling from a categorical distribution is linear in the number of
categories \cite{Devroye86}. Consequently, the complexity of the Gibbs
sampler (Algorithm 1) is $\mathcal{O}(TP^{2}M)$.

Proposition~\ref{marg_cond} also implies that for a given a positive 1-1 $%
\gamma _{\bar{n}}$, only $\gamma _{n}\in \{-1$:$M\}$ that does not violate
the positive \mbox{1-1} property can be generated by the conditional $\pi
_{n}(\cdot |\gamma _{\bar{n}})$, with probability proportional to $\eta
_{n}(\gamma _{n})$. Thus, starting with a positive 1-1 vector, all iterates
of the Gibbs sampler are also positive 1-1. If the chain is run long enough,
the samples are effectively distributed from (\ref{eq:thetajoint_dis}) as
formalized in Proposition \ref{convergence} (see Appendix for proof).

\begin{proposition}
\label{convergence} Starting from any initial state in ${\Gamma }$, the
Gibbs sampler defined by the family of conditionals (\ref{eq:marg_cond1})
converges to the target distribution (\ref{eq:thetajoint_dis}) at an
exponential rate. More concisely, let $\pi ^{j}$ denote the $j$th power of
the transition matrix, then
\begin{equation*}
\max_{\gamma ,\gamma ^{\prime }\in {\Gamma }}(|\pi ^{j}(\gamma ^{\prime
}|\gamma )-\pi (\gamma ^{\prime })|)\leq (1-2\beta )^{\left\lfloor \frac{j}{2%
}\right\rfloor },
\end{equation*}%
where $\beta \triangleq \min_{\gamma ,\gamma ^{\prime }\in {\Gamma }}\pi
^{2}(\gamma ^{\prime }|\gamma )>0$ is the least likely 2-step transition
probability.
\end{proposition}

\vspace{5pt}

\hrule

\vspace{5pt}

\textbf{Algorithm 1a. }$\gamma _{n}^{\prime }\sim \pi _{n}(\cdot |\gamma
_{1:n-1}^{\prime },\gamma _{n+1:P})$

\vspace{5pt}

\hrule

\vspace{5pt}

$c:=[-1:M];\quad \eta _{n}:=[\eta _{n}(-1),...,\eta _{n}(M)];$

\textsf{{\footnotesize {for }}}$j=1:M$

\quad $\eta _{n}(j):=\eta _{n}(j)(1-1_{\{\gamma _{1:n-1}^{\prime },\gamma
_{n+1:P}\}}(j));$

\textsf{{\footnotesize {end}}}

$\gamma _{n}^{\prime }\sim \mathsf{Categorical}(c,\eta _{n});$

\vspace{4pt}

\hrule

\vspace{8pt}

The proposed Gibbs sampler has a short burn-in period due to its exponential
convergence rate. More importantly, since we are not using the samples to
approximate (\ref{eq:thetajoint_dis}) as in an MCMC inference problem, it is
not necessary to discard burn-in and wait for samples from the stationary
distribution. For the purpose of approximating the $\delta $-GLMB filtering
density, each distinct sample constitutes one term in the approximant, and
reduces the $L_{1}$ approximation error by an amount proportional to its
weight. Hence, regardless of their distribution, all distinct samples can be
used, the larger the weights, the smaller the $L_{1}$ error between the
approximant and the true $\delta $-GLMB.

\begin{remark}
The proposed Gibbs sampling solution can be directly applied to the standard
data association problem in joint probabilistic data association and
multiple hypothesis tracking. Further, it can be adapted to solve a ranked
assignment problem with $P$ workers and $M$ jobs as follows. Each assignment
is represented by a positive 1-1 vector $\gamma $ in $\{0$:$M\}^{P}$, with $%
\gamma _{n}=j$ indicating that worker $n$ is assigned job $j$, which incurs
a cost $F_{n,j}$. Note that $\gamma _{n}=0$ indicates that worker $n$ is
assigned no job, which incurs a cost of $F_{n,0}$ (usually assumed to be 0).
The cost of an assignment $\gamma $ is given by $\tsum\nolimits_{n=1%
\!}^{P}F_{n,\gamma _{n}}$. Hence, Algorithm 1 can be used to generate a
sequence of assignments with significant costs by setting $c:=[0:M]$, and $%
\eta _{n}=[\eta _{n}(0),...,\eta _{n}(M)]$, where $\eta _{n}(j)=\exp
(-F_{n,j})$, in the first line of Algorithm 1a. A sufficient condition for
exponential convergence of the Gibbs sampler (i.e. Proposition \ref%
{convergence} to hold) is a finite cost of assigning no job to any worker
(usually this cost is $0$, i.e. $\eta _{i}(0)=1$). The final step is to
remove duplicates and rank the Gibbs samples (according to their costs),
which requires additional computations with $\mathcal{O}(T\log T)$
complexity.
\end{remark}

\subsection{Joint Prediction and Update Implementation\label%
{subsec:implementation}}

A $\delta $-GLMB of the form (\ref{eq:delta-glmb}) is completely
characterized by parameters $(\omega ^{(I,\xi )},p^{(\xi )})$, $(I,\xi )\in
\mathcal{F}\!(\mathbb{L})\!\times \!\Xi $, which can be enumerated as $%
\{(I^{(h)},\xi ^{(h)},\omega ^{(h)},p^{(h)})\}_{h=1}^{H}$, where%
\begin{equation*}
\omega ^{(h)}\triangleq \omega ^{(I^{(h)},\xi ^{(h)})},\;p^{(h)}\triangleq
p^{(\xi ^{(h)})}.
\end{equation*}%
Since the $\delta $-GLMB (\ref{eq:delta-glmb}) can now be rewritten as
\begin{equation*}
\mathbf{\pi }(\mathbf{X})=\Delta (\mathbf{X})\sum\limits_{h=1}^{H}\omega
^{(h)}\delta _{I^{(h)}}[\mathcal{L(}\mathbf{X})]\left[ p^{(h)}\right] ^{%
\mathbf{X}},
\end{equation*}%
there is no need to store $\xi ^{(h)}$, and implementing the GLMB filter
amounts to propagating forward the parameter set
\begin{equation*}
\{(I^{(h)},\omega ^{(h)},p^{(h)})\}_{h=1}^{H}.
\end{equation*}%
Estimating the multi-object state from the $\delta $-GLMB parameters is the
same as in \cite{VVP_GLMB13}.

The procedure for computing the parameter set
\begin{equation*}
\{(I_{+}^{(h_{+})},\omega _{+}^{(h_{+})},p_{+}^{(h_{+})})\}_{h_{+}=1}^{H_{+}}
\end{equation*}%
at the next time is summarized in Algorithm~2. Note that to be consistent
with the indexing by $h$ instead of $(I,\xi )$, we abbreviate%
\begin{eqnarray}
\!\!\!\!\!\!\!\bar{P}_{S}^{(h)}(\ell _{_{\!}i})\!\! &\triangleq &\!\!\bar{P}%
_{_{\!}S\!}^{(\xi ^{(h)})}(\ell _{_{\!}i}),  \notag \\
\!\!\!\!\!\!\!\bar{p}_{+}^{(h)_{\!}}(x,\ell _{_{\!}i})\!\! &\triangleq &\!\!%
\bar{p}_{+}^{(\xi ^{(h)})_{\!}}(x,\ell _{_{\!}i\!}),  \notag \\
\!\!\!\!\!\!\!\bar{\psi}_{Z_{+}}^{(h,j)}(\ell _{_{\!}i})\!\! &\triangleq
&\!\!\bar{\psi}_{Z_{+}}^{(\xi ^{(h)},j)}(\ell _{_{\!}i})  \notag \\
\!\!\!\!\!\!\!\eta _{i}^{(h)}(j)\!\! &\triangleq &\!\!%
\begin{cases}
1-\bar{P}_{S}^{(h)\!}(\ell _{i}), & \!\ell _{i\!}\in I^{(h)},\text{ }j\!<0\!,
\\
\bar{P}_{S}^{(h)}(\ell _{i})\bar{\psi}_{Z_{+}}^{(h,j)\!}(\ell _{i\!}), &
\!\ell _{i\!}\in I^{(h)},\text{ }j\!\geq 0, \\
1-r_{B\!,+}(\ell _{i}), & \!\ell _{i\!}\in \mathbb{B}_{+},\text{ }j\!<0, \\
r_{B\!,+}(\ell _{i})\bar{\psi}_{Z_{+}}^{(h,j)\!}(\ell _{i}), & \!\ell
_{i\!}\in \mathbb{B}_{+},\text{ }j\!\geq 0.%
\end{cases}
\label{eq:eta_h}
\end{eqnarray}%
At time $k+1$, the three main tasks are:

\begin{enumerate}
\item Generate $\{(I^{(h_{+})},\xi ^{(h_{+})},I_{+}^{(h_{+})},\theta
_{+}^{(h_{+})})\}_{h_{+}=1}^{H_{+}}$, the\ set of \textquotedblleft
children\textquotedblright\ with significant weights;

\item Compute $\{(I^{(h_{+})},I_{+}^{(h_{+})},\omega
_{+}^{(h_{+})},p_{+}^{(h_{+})})\}_{h_{+}=1}^{H_{+}}$, the intermediate
parameter set, as in Proposition \ref{Prop_joint};

\item Compute $\{(I_{+}^{(h_{+})},\omega
_{+}^{(h_{+})},p_{+}^{(h_{+})})\}_{h_{+}=1}^{H_{+}}$, the parameter set at
time $k+1$, using (\ref{eq:GLMB_joint6}).
\end{enumerate}

For task 1, using the rationale from subsection \ref{subsec:fast_assginment}%
, the set of \textquotedblleft children\textquotedblright\ can be generated
by sampling from the distribution $\pi $ given by
\begin{equation*}
\pi (I,\xi ,I_{+},\theta _{+})\propto \omega ^{(I,\xi )}\omega
_{Z_{_{\!}+}}^{(I,\xi ,I_{+},\theta _{+})}.
\end{equation*}%
This can be achieved by sampling $(I^{(h_{+})},\xi ^{(h_{+})})$ from $\pi
(I,\xi )\propto \omega ^{(I,\xi )}$, and then conditional on $%
(I^{(h_{+})},\xi ^{(h_{+})})$, sample $(I_{+}^{(h_{+})},\theta
_{+}^{(h_{+})})$\ from $\pi (I_{+},\theta _{+}|I^{(h_{+})},\xi ^{(h_{+})})$.
Equivalently, in Algorithm~2 we draw $H_{+}^{\max }$ samples $(I^{(h)},\xi
^{(h)})$ from $\pi (I,\xi )\propto \omega ^{(I,\xi )}$, and then for each
distinct sample $(I^{(h)},\xi ^{(h)})$ with $T_{+}^{(h)}$ copies\footnote{%
Asymptotically $T_{+}^{(h)}$ is proportional to the weight $\omega ^{(h)}$},
use the Gibbs sampler (Algorithm 2a) to generate $T_{+}^{(h)}$ samples $%
(I_{+}^{(h,t)},\theta _{+}^{(h,t)})$ from $\pi (I_{+},\theta
_{+}|I^{(h)},\xi ^{(h)})$. Note that each $(I_{+}^{(h,t)},\theta
_{+}^{(h,t)})$ is represented by the positive 1-1 vector $\gamma ^{(h,t)}$
(see (\ref{eq:convert}) for the equivalence of this representation).

\bigskip

\hrule

\vspace{5pt}

\textbf{Algorithm 2. Joint Prediction and Update}\footnote{%
In Algorithm 2 $\{\}$ denotes a MATLAB\ cell array of (non-unique) elements}

\begin{itemize}
\item {\footnotesize \textsf{input: }}$\{(I^{(h)},\omega
^{(h)},p^{(h)})\}_{h=1}^{H}$, $Z_{+}$, $H_{+}^{\max }$,

\item {\footnotesize \textsf{input: }}$\{(r_{\!B\!,+}^{(\ell
)},p_{B\!,+}^{(\ell )})\}_{\ell \in \mathbb{B}_{\!+}}$, $P_{S}$, $%
f_{\!+\!}(\cdot |\cdot )$, $\kappa _{+}$, $P_{D\!,+}$, $g_{+\!}(\cdot |\cdot
)$,

\item \textsf{{\footnotesize {output: }}}$\{(I_{+}^{(h_{+})},\omega
_{+}^{(h_{+})},p_{+}^{(h_{+})})\}_{h_{+}=1}^{H_{+}}$
\end{itemize}

\vspace{2pt}

\hrule

\vspace{2pt}

\textsf{{\footnotesize {sample counts }}}$[T_{+}^{(h)}]_{h=1}^{H}$\textsf{%
{\footnotesize {\ \ from\textsf{\ a multinomial} distribution with}}}

\textsf{{\footnotesize {parameters }}}$H_{+}^{\max }$\textsf{{\footnotesize {%
\ trials and weights }}}${[}\omega ^{(h)}]_{h=1}^{H}$

\textsf{{\footnotesize {for}}}\textsf{\ }$h=1:H$

\quad \textsf{{\footnotesize {initialize }}}$\gamma ^{(h,1)}$

\quad \textsf{{\footnotesize {compute }}}$\eta ^{(h)}=[\eta
_{i}^{(h)}(j)]_{(i,j)=(1,-1)}^{(|I^{(h)}\!\cup \!\mathbb{B}_{+}|,|Z_{+}|)}$
\ \textsf{{\footnotesize {using }}}(\ref{eq:eta_h})

\quad $\{\gamma ^{(h,t)}\}_{_{t=1}}^{\tilde{T}_{+}^{(h)}}:=\mathsf{%
Unique(Gibbs}(\gamma ^{(h,1)},T_{+}^{(h)},\eta ^{(h)}));$

\quad \textsf{{\footnotesize {for }}}$t=1:\tilde{T}_{+}^{(h)}$

\quad \quad \textsf{{\footnotesize {compute }}}$I_{+}^{(h,t)}$\textsf{%
{\footnotesize {\ from }}}$I^{(h)}$\textsf{{\footnotesize {\ and }}}$\gamma
^{(h,t)}$\textsf{{\footnotesize {\ using }}}(\ref{eq:Iplus})

\quad \quad \textsf{{\footnotesize {compute }}}$\omega _{+}^{(h,t)}$\textsf{%
{\footnotesize {\ from }}}$\omega ^{(h)}$\textsf{{\footnotesize {\ and }}}$%
\gamma ^{(h,t)}$\textsf{{\footnotesize {\ using }}}(\ref{eq:wplus})

\quad \quad \textsf{{\footnotesize {compute }}}$p_{+}^{(h,t)}$\ \textsf{%
{\footnotesize {from }}}$p^{(h)}$\textsf{{\footnotesize {\ and }}}$\gamma
^{(h,t)}$\textsf{{\footnotesize {\ using}}} (\ref{eq:pplus})

\quad \textsf{{\footnotesize {end}}}

\textsf{{\footnotesize {end}}}

$(\{(I_{+}^{(h_{+})}\!,p_{+}^{(h_{+})})\}_{h_{+}=1}^{H_{+}},\sim ,[U_{h,t}])$

\quad \quad \quad \quad \quad \quad \quad \quad $:=\mathsf{Unique}%
(\{(I_{+}^{(h,t)}\!,p_{+}^{(h,t)})\}_{(h,t)=(1,1)}^{(H,\tilde{T}%
_{+}^{(h)})});$

\textsf{{\footnotesize {for }}}$h_{+}=1:H_{+}$

\quad $\omega _{+}^{(h_{+})}:=\sum\limits_{h,t:U_{h,t}=h_{+}}\omega
_{+}^{(h,t)};$

\textsf{{\footnotesize {end}}}

\textsf{{\footnotesize {normalize weights }}}$\{\omega
_{+}^{(h_{+})}\}_{h_{+}=1}^{H_{+}}$

\vspace{4pt}

\hrule

\vspace{5pt}

\hrule

\vspace{3pt}

\vspace{5pt}

\textbf{Algorithm 2a. Gibbs }

\begin{itemize}
\item \textsf{{\footnotesize {input: }}}$\gamma ^{(1)},T,\eta =[\eta
_{i}(j)] $

\item \textsf{{\footnotesize {output: }}}$\gamma ^{(1)},...,\gamma ^{(T)}$
\end{itemize}

\vspace{2pt}

\hrule

\vspace{4pt}

$P:=\mathsf{size}(\eta ,1);\ M:=\mathsf{size}(\eta ,2)-2;\ c:=[-1$:$M];\
\tilde{\eta}:=\eta ;$

\textsf{{\footnotesize {for}}\ }$t=2:T$

\quad $\gamma ^{(t)}:=[$ $];$

\quad \textsf{{\footnotesize {for }}}$n=1:P$

\quad \quad \textsf{{\footnotesize {for }}}$j=1:M$

\quad \quad \quad $\tilde{\eta}_{n}(j):=\eta _{n}(j)(1-1_{\{\gamma
_{1:n-1}^{(t)},\gamma _{n+1:P}^{(t-1)}\}}(j));$

\quad \quad \textsf{{\footnotesize {end}}}

\quad \quad $\gamma _{n}^{(t)}\sim \mathsf{Categorical}(c,\tilde{\eta}_{n});$%
\quad $\gamma ^{(t)}:=[\gamma ^{(t)},\gamma _{n}^{(t)}];$

\quad \textsf{{\footnotesize {end}}}

\textsf{{\footnotesize {end}}}

\vspace{5pt}

\hrule

\bigskip

In task 2, for each $h$, after discarding repeated positive \mbox{1-1}
vector samples via the \textquotedblleft Unique\textquotedblright\ MATLAB
function, the intermediate parameters $(I^{(h)},I_{+}^{(h,t)},\omega
_{+}^{(h,t)},p_{+}^{(h,t)})$, $t=1$:$\tilde{T}_{+}^{(h)}$ are computed from
the positive 1-1 vector $\gamma ^{(h,t)}$ by
\begin{eqnarray}
\!\!\!\!I_{+}^{(h,t)}\!\!\!\! &=&\!\!\!\!\{\ell _{i}\in I^{(h)}\cup \mathbb{B%
}_{\!{+}\!}:\gamma _{i}^{(h,t)}\geq 0\},  \label{eq:Iplus} \\
\!\!\!\!\omega _{+}^{(h,t)}\!\!\!\! &\propto &\!\!\!\!\omega
^{(h)}\tprod\limits_{i=1}^{|I^{(h)}\cup \mathbb{B}_{+}|}\eta
_{i}^{(h)}(\gamma _{i}^{(h,t)}),  \label{eq:wplus} \\
\!\!\!\!p_{+}^{(h,t)\!}(\cdot ,\ell _{i})\!\!\!\! &=&\!\!\!\!\bar{p}%
_{+}^{(h)}(\cdot ,\ell _{i})\psi _{Z_{+}}^{(\gamma _{i}^{(h,t)})}(\cdot
,\ell _{i})/\bar{\psi}_{Z_{+}}^{(h,\gamma _{i}^{(h,t)})}(\ell _{i}).
\label{eq:pplus}
\end{eqnarray}%
Equations (\ref{eq:Iplus})-(\ref{eq:pplus}) follow directly from (\ref%
{eq:convert}), (\ref{eq:GLMB_joint1})-(\ref{eq:GLMB_joint5}) and (\ref%
{eq:eta_prod}). Computing $p_{+}^{(h,t)}(\cdot ,\ell _{i})$ (and $\eta
_{i}^{(h,t)}(j)$, $\bar{\psi}_{Z_{+}}^{(h,j)}(\ell _{i})$, $\bar{P}%
_{_{\!}S\!}^{(h)\!}$)\ can be done via sequential Monte Carlo or Gaussian
mixture, depending on the representation of the densities $p^{(h)}$ (see
subsections IV.B and V.B of \cite{VVP_GLMB13} for details).

Finally, in task 3 the intermediate parameters are marginalized via (\ref%
{eq:GLMB_joint6}) and the weights normalized, to give the new component set $%
\{(I_{+}^{(h_{+})},\omega
_{+}^{(h_{+})},p_{+}^{(h_{+})})\}_{h_{+}=1}^{H_{+}} $. Note that the output $%
U_{h,t}$ of the \textquotedblleft Unique\textquotedblright\ MATLAB function
gives the index of the GLMB component at time $k+1$ that $%
(I^{(h)},I_{+}^{(h,t)},p_{+}^{(h,t)})$ contributes to.

Since we are only interested in samples that provide a good representation
of the $\delta $-GLMB filtering density, increased efficiency (for the same $%
H_{+}^{\max }$) can be achieved by using annealing or tempering techniques
to modify the stationary distribution so as to induce the Gibbs sampler to
seek more diverse samples \cite{GeyerThompson95}, \cite{Neal2000}. One
example is to initially decrease the temperature to seek out the nearest
mode, and subsequently increase the temperature for diversity.

In scenarios with small birth weights, $\delta $-GLMB components that
involve births also have small weights, and are likely to be discarded when $%
H_{+}^{\max }$ (which depends on the available computing resource) is not
large enough, leading to poor track initiation. Increasing the temperature
does not guarantee the selection of components with births. Tempering with
the birth model (e.g. by feeding the Gibbs sampler with a larger birth rate)
directly induces the chain to generate more components with births. Note
that the actual weights of the $\delta $-GLMB density components are
computed using the correct birth model parameters. Similarly, tempering with
the survival probability induces the Gibbs sampler to generate more
components with object deaths and improves track termination. Tempering with
parameters such as detection probabilities and clutter rate induces the
Gibbs sampler to generate components that reduce the occurrence of dropped
tracks. Note that if the number of significant components exceeds $%
H_{+}^{\max }$ the filtering performance will degrade in subsequent
iterations.

The Gibbs sampler can be initialized with the highest weighted 1-1 vector
(requires solving an optimal assignment problem \cite{Munkres57}, \cite%
{Jonker87}). Alternatively, a trivial initialization is the all-zeros 1-1
vector (requires no computations). Proposition \ref{convergence} ensures
convergence of the chain to the stationary distribution at an exponential
rate regardless of the initialization.

\begin{remark}
It is possible to replace the Gibbs sampler by a deterministic ranked
assignment algorithm with cost matrix (\ref{eq:Ass_Cost_Matrix1}) to
generate the $T_{+}^{(h)}$ strongest positive 1-1 vectors, where $%
T_{+}^{(h)} $ is chosen to be proportional to the weight $\omega ^{(h)}$
\cite{HVV15}. Note that such allocation scheme does not necessarily produce
the $H_{+}$ best components at time $k+1$. It is possible to discard some
children from weaker parents, which still have higher weights than some of
those from stronger parents. Other schemes for choosing $T_{+}^{(h)}$ are
possible. A ranked assignment problem can also be formulated to find the $%
H_{+}$ best components. However, the complexity grows and parallelizability
is lost. Similar to the Gibbs sampler based solution, tempering with the
multi-object model parameters can be used to increase efficiency.
\end{remark}

Let $P\triangleq \max_{h}|I^{(h)}\!\cup \!\mathbb{B}|$ and $M\triangleq
|Z_{+}|$. The standard and fastest ranked assignment algorithms have
respective complexities $\mathcal{O}((2P+M)^{4})$ and $\mathcal{O}%
((2P+M)^{3})$ i.e. at best, cubic in both the number of hypothesized labels
and measurements. On the other hand, the complexity of the Gibbs sampling
based solution is $\mathcal{O}(P^{2}M)$, i.e. quadratic in the number of
hypothesized labels and linear in the number of measurements.

\section{Numerical Studies\label{sec:sim}}

This section presents two numerical experiments to verify the proposed GLMB
filter implementation without consideration for parallelization. The first
demonstrates its efficiency via a linear Gaussian scenario. The second
demonstrates its versatility on a very challenging non-linear scenario with
non-uniform detection profile and dense clutter.

\subsection{Linear Gaussian}

The linear Gaussian scenario in the experiment of \cite{VVP_GLMB13} is used
to compare typical speedup in CPU time between the original implementation
in \cite{VVP_GLMB13} and the proposed implementation (Algorithm 2). In
summary this scenario involves an unknown and time varying number objects
(up to 10 in total) with births, deaths and crossings. Individual object
kinematics are described by a 4D state vector\ of position and velocity that
follows a constant velocity model with sampling period of\ $1s$, and process
noise standard deviation $\sigma _{\nu }=5m/s^{2}$. The survival probability
$P_{S}=0.99$, and the birth model is an LMB with parameters $\{r_{B,k}(\ell
_{i}),p_{B,k}(\ell _{i})\}_{i=1}^{3}$, where $\ell _{i}=(k,i)$, $%
r_{B,k}(\ell _{i})=0.04$, and $p_{B}(x,\ell _{i})=\mathcal{N}%
(x;m_{B}^{(i)},P_{B})$ with%
\begin{equation*}
\begin{array}{ll}
m_{B}^{(1)}=[0,0,100,0]^{T}, & \!\!m_{B}^{(2)}=[-100,0,-100,0]^{T}, \\
m_{B}^{(3)}=[100,0,-100,0]^{T}, & \!\!P_{B}^{\text{ \ \ \ }}=\mathrm{diag}%
([10,10,10,10]^{T})^{2}.%
\end{array}%
\end{equation*}

Observations are 2D position vectors on the region $[-1000,1000]m\times
\lbrack -1000,1000]m$ with noise standard deviation $\sigma _{\varepsilon
}=10m$. The detection probability $P_{D}=0.88$\ and clutter is modeled as a
Poisson RFS with a uniform intensity of $\lambda _{c}=1.65\times
10^{-5}~m^{-2}$ on the observation region (i.e. an average of 66 false
alarms per scan).

In the original implementation, predicted components are obtained by
combining independently generated surviving and births components, together
with a CPHD look ahead step for better efficiency (see \cite{VVP_GLMB13} for
full details). To increase diversity of the GLMB components, the proposed
implementation with Gibbs sampling uses tempered birth, survival and
detection parameters, specifically each $r_{B}(_{\!}\ell _{i\!})$ is
increased by a factor of 10, while $P_{S}$ and $P_{D}$ are reduced by 5\%.
For completeness, the implementation via joint prediction and update with
Murty's algorithm is also considered using the same tempered parameters. All
implementations use Gaussian mixture representations of the track densities.
We compare the speedup (ratio of CPU times) between the original
implementation and the proposed Gibbs sampler based joint prediction and
update implementation as well as between the original implementation and the
Murty based joint prediction and update implementation. Three cases are
considered for various values of $H^{\max }$ (the cap on the number of
components) under which the implementations exhibit approximately the same
tracking performance (it is virtually impossible for the different
implementations to have exactly the same tracking performance). The results
reported below are obtained over 100 Monte Carlo trials.

\begin{figure}[h]
\begin{center}
\resizebox{80mm}{!}{\includegraphics{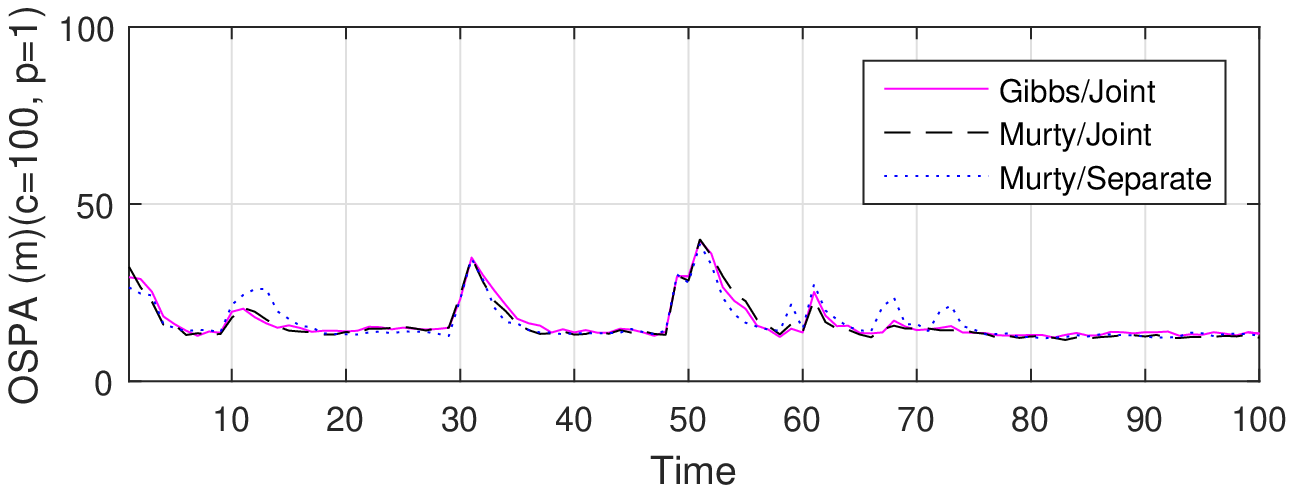}}
\end{center}
\caption{Similar OSPA errors for the three different implementations. }
\label{fig:lgospa}
\end{figure}

\begin{figure}[h]
\begin{center}
\begin{tabular}{c|ccc}
\hline\hline
\textit{\textbf{Speedup}} & \textbf{Case 1} & \textbf{Case 2} & \textbf{Case
3} \\ \hline
\textbf{Murty/Joint} & \textit{24X} & \textit{2.5X} & \textit{185X} \\ \hline
\textbf{Gibbs/Joint} & \textit{187X} & \textit{27X} & \textit{1443X} \\
\hline\hline
\end{tabular}%
\end{center}
\par
\vspace{0.2cm}
\caption{Range of CPU time speedup factors compared to the original
implementation obtained for various cases.}
\label{fig:speedup}
\end{figure}

\subsubsection{Case 1}

This baseline comparison uses the values of $H^{\max }$ where each
implementation starts to exhibit reasonable tracking performance, and have
approximately the same average optimal sub-pattern assignment (OSPA) error
\cite{SVV08}. The original implementation requires $H^{\max }=10^{4}$ while
the joint prediction and update implementation with the proposed Gibbs
sampler and Murty's algorithm both require $H^{\max }=10^{3}$. Fig. \ref%
{fig:lgospa} confirms that all implementations exhibit approximately similar
OSPA curves, except for several pronounced peaks between times $k=55$ and $%
k=75$ for the original implementation, due to the latter being slower
confirm new births. Fig. \ref{fig:speedup} shows speedups of about two and
one orders of magnitude, respectively, for the proposed Gibbs based and
Murty based joint prediction and update implementations.

\subsubsection{Case 2}

All implementations are allocated the same $H^{\max }=10^{4}$. Fig. \ref%
{fig:speedup} shows a speedup of over one order of magnitude for the
proposed joint prediction and update implementation with Gibbs sampling
while there is a small improvement in the Murty based joint prediction and
update implementation. Both joint prediction and update implementations now
only show a slightly better average OSPA error than the original
implementation, confirming that $H^{\max }=10^{3}$ is a good trade-off
between computational load and performance.

\subsubsection{Case 3}

In an attempt to reduce the peaks in the OSPA curve observed in case 1, we
raise $H^{\max }$ for the original implementation to $10^{5}$ (the
experiment takes too long to run for larger values of $H^{\max }$ to be
useful). However, these peaks only reduce slightly and are still worse than
both joint prediction and update implementations for $H^{\max }=10^{3}$.
Furthermore, the original implementation now only shows a slightly better
average OSPA error than the others. In this extreme case Fig. \ref%
{fig:speedup} shows speedups of roughly three and two orders of magnitude,
respectively, for the Gibbs based and Murty based joint prediction and
update implementations.

It should be noted that in general the actual speedup observed depends
strongly on the scenario and testing platform, and hence the reported
speedup figures should be taken only as broad indication of the range that
could be expected. For an indication of the actual speed and accuracy, on
real data, against some recent algorithms, we refer the reader to \cite%
{KVV16}.

\subsection{Non-linear}

This example considers a very challenging scenario in which previous
implementations breakdown, specifically the non-linear scenario in the
experiment of \cite{VoGLMB13}, with reduced detection profile and increased
clutter rate. Again there is an unknown and time varying number of objects
(up to 10 in total) with births, deaths, and crossings. Individual object
kinematics are described by a 5D state vector $x_{k}=[~p_{x,k},\dot{p}%
_{x,k},p_{y,k},\dot{p}_{y,k}~,\omega _{k}]^{T}$ of planar position,
velocity, and turn rate, which follows a coordinated turn model with a
sampling period of $1s$ and transition density $f_{k|k-1}(x_{k}|x_{k-1})=%
\mathcal{N}(x_{k};F(\omega _{k})x_{k},Q)${,} where%
\begin{equation*}
F(\omega )=\left[
\begin{array}{ccccc}
1 & \!\!\frac{\sin \omega }{\omega } & 0 & \!\!-\frac{1-\cos \omega }{\omega
} & 0 \\
0 & \!\!\cos \omega & 0 & \!\!-\sin \omega & 0 \\
0 & \!\!\frac{1-\cos \omega }{\omega } & 1 & \!\!\frac{\sin \omega }{\omega }
& 0 \\
0 & \!\!\sin \omega & 0 & \!\!\cos \omega & 0 \\
0 & 0 & 0 & 0 & 1%
\end{array}%
\right] \!\!,G=%
\begin{bmatrix}
\frac{1}{2} & 0 \\
1 & 0 \\
0 & \frac{1}{2} \\
0 & 1%
\end{bmatrix}%
\!\!,
\end{equation*}%
\ $Q=\mathrm{diag}([\sigma _{w}^{2}GG^{T},\sigma _{u}^{2}])$, $\sigma
_{w}=15m/s^{2}$, and $\sigma _{u}=(\pi /180)rad/s$ are the process noise
standard deviations. The survival probability $P_{S}=0.99$, and the birth
model is an LMB with parameters $\{r_{B,k}(\ell _{i}),p_{B,k}(\ell
_{i})\}_{i=1}^{4}$, where $\ell _{i}=(k,i)$, $r_{B,k}(\ell
_{1})=r_{B,k}(\ell _{2})=0.02$, $r_{B,k}(\ell _{3})=r_{B,k}(\ell _{4})=0.03$%
, and $p_{B,k}(x,\ell _{i})=\mathcal{N}(x;m_{B}^{(i)},P_{B})$ with%
\begin{eqnarray*}
&&\!\!\!\!\!\!\!\!\!\!\!\!\!\!%
\begin{array}{ll}
m_{B}^{(1)}=[-1500,0,250,0,0]^{T}, & \!\!m_{B}^{(2)}=[-250,0,1000,0,0]^{T}
\\
m_{B}^{(3)}=[250,0,750,0,0]^{T}, & \!\!m_{B}^{(4)}=[1000,0,1500,0,0]^{T},%
\end{array}
\\
&&\!\!\!\!\!\!\!\!\!\!\!\!\!\!%
\begin{array}{ll}
P_{B}^{\text{ \ \ \ }}=\mathrm{diag}([50,50,50,50,6(\pi /180)]^{T})^{2}. &
\end{array}%
\end{eqnarray*}

Observations are noisy 2D bearings and range detections $z=[~\theta ,r~]^{T}$
on the half disc of radius $2000m$ with noise standard deviations $\sigma
_{r}=5m$ and $\sigma _{\theta }=(\pi /180)rad$ respectively. The detection
profile is a (unnormalized) Gaussian with a peak of $0.95$ at the origin and
$0.88$ at the edge of surveillance region. Clutter follows a Poisson RFS
with a uniform intensity of $\lambda _{c}=1.6\times 10^{-2}$ $(radm)^{-1}$
on the observation region (i.e. an average of 100 false alarms per scan).

\begin{figure}[h]
\begin{center}
\resizebox{80mm}{!}{\includegraphics{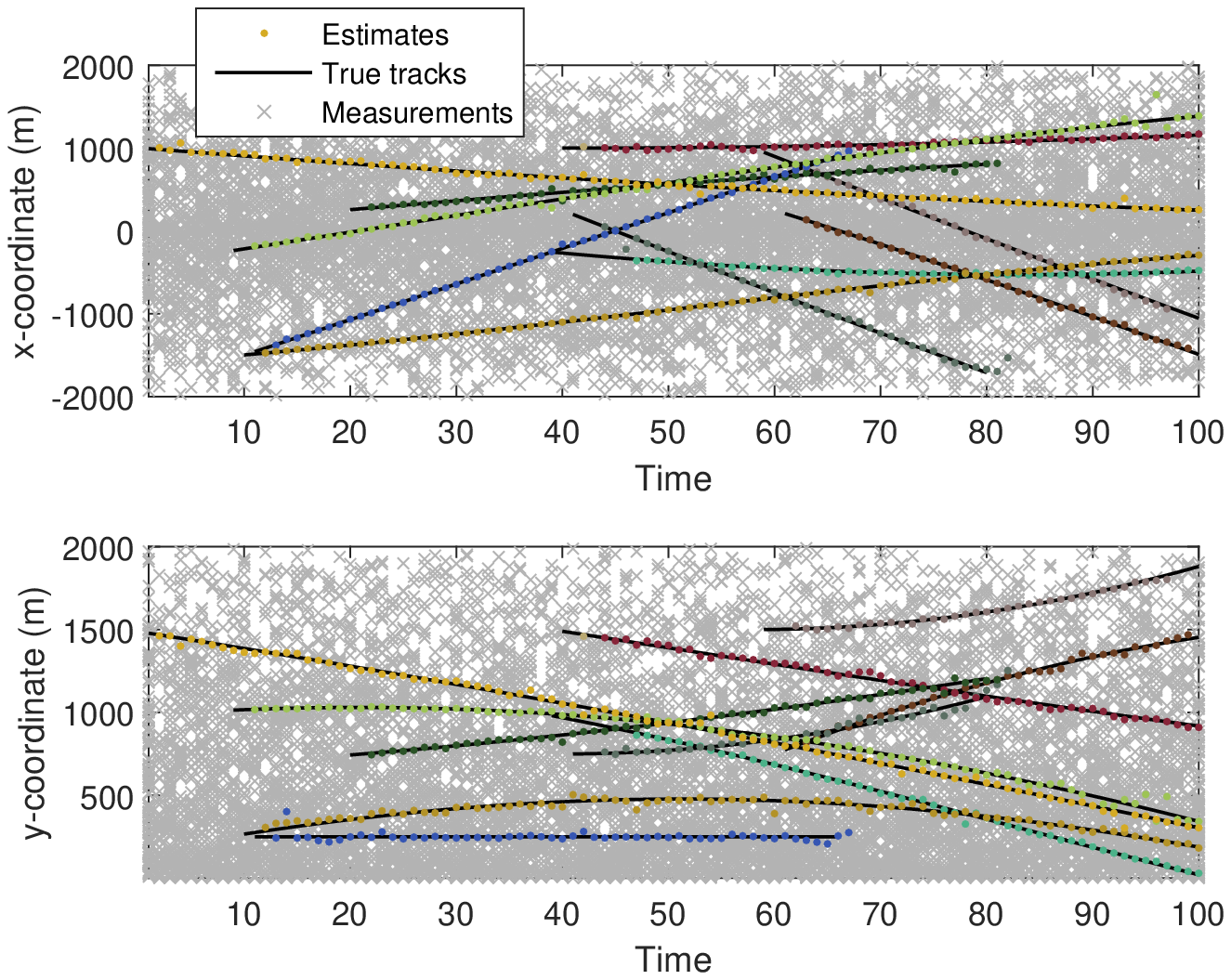}}
\end{center}
\caption{Track estimates, measurements and ground truths for a sample run.}
\label{fig:smcxyt}
\end{figure}

\begin{figure}[h]
\begin{center}
\resizebox{80mm}{!}{\includegraphics{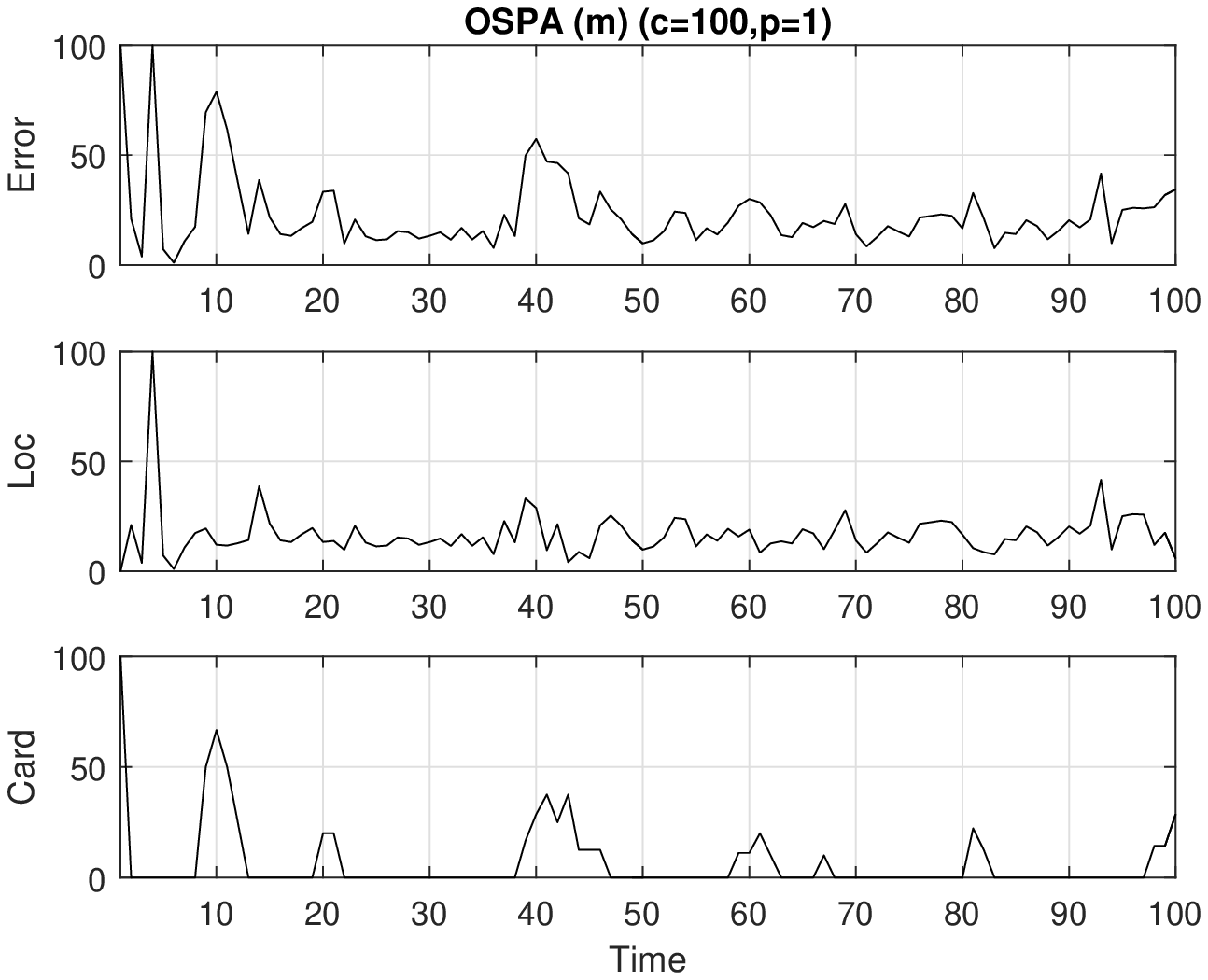}}
\end{center}
\caption{OSPA error for a sample run. }
\label{fig:smcospa}
\end{figure}

The proposed GLMB filter implementation uses particle approximations of the
track densities to accommodate nonlinearity and state dependent probability
of detection. Again, the birth, survival and detection parameters are
tempered, specifically each $r_{B}(\ell _{i\!})$ is increased by a factor of
20, while $P_{S}(x,\ell )$ and $P_{D}(x,\ell )$ are reduced by 5\%. Due to
the high uncertainty in the scenario, a large $H^{\max }$ is needed since
there is a large number of GLMB components with similar weights. Further,
the differences between the weights of significant and insignificant
components are not as pronounced as the original scenario in \cite{VoGLMB13}%
. Consequently, the Monte Carlo integration error should be significantly
smaller than these differences for the computation of (\ref{eq:eta_h}) and
the weights to be useful. Thus regardless of whether Gibbs sampling or
Murty's algorithm is used, a very large number of particles per object is
needed. The tracking result with $H^{\max }=10^{5}$ and 100000 particles per
object for a single sample run shown in Fig. \ref{fig:smcxyt} illustrates
that the proposed GLMB filter implementation successfully track all objects,
and is confirmed by the OSPA\ curve in Fig. \ref{fig:smcospa}. Previous
implementations break down due to the large number of required components
and particles.

\section{Conclusions}

\label{sec:sum} This paper proposed an efficient implementation of the GLMB
filter by integrating the prediction and update into one step along with an
efficient algorithm for truncating the GLMB filtering density based on Gibbs
sampling. The resulting algorithm is an on-line multi-object tracker with
linear complexity in the number of measurements and quadratic in the number
of hypothesized tracks, which can accommodate non-linear dynamics and
measurements, non-uniform survival probabilities, sensor field of view and
clutter intensity. This implementation is also applicable to approximations
such as the labeled multi-Bernoulli (LMB) filter since this filter requires
a special case of the $\delta $-GLMB prediction and a full $\delta $-GLMB
update to be performed \cite{ReuterLMB14}. The proposed Gibbs sampler can be
adapted to solve the ranked assignment problem and hence the data
association problem in other tracking approaches. It is also possible to
parallelize the Gibbs sampler \cite{ParalleGibbs_11}. A venue for further
research is the generalization of the proposed technique to more complex
problems such as multiple extended object tracking \cite{BVV15}, or tracking
with merged measurements \cite{Beard_etal16}. 

\section{Appendix: Mathematical Proofs}

\textbf{Proof of Proposition \ref{Prop_joint}:} Using the change of variable
$I_{+\!}=J\cup L_{+}$, we have $J=\mathbb{L}\cap I_{+}$, $L_{+}=\mathbb{B}%
_{\!{+}\!}\cap I_{+}$, and hence (\ref{eq:GLMB_upd1}) becomes%
\begin{eqnarray*}
\omega _{Z_{_{\!}+}}^{\!(J,L_{\!+\!,}\xi ,\theta _{\!+\!})}\!\!\!
&=&\!\!\!\omega _{Z_{_{\!}+}}^{(\mathbb{L}\cap I_{+},\mathbb{B}_{{+}}\cap
I_{+}\xi ,\theta _{\!+})} \\
\!\!\! &=&\!\!\!1_{_{\!}{\Theta }_{\!+\!}(I_{+})}(\theta _{\!+\!})\!\left[
_{\!}\bar{\psi}_{_{\!}Z_{_{\!}+}}^{(\xi ,\theta _{_{\!}+})\!}\right]
^{\!I_{+}}\!\bar{\omega}_{+}^{(\mathbb{L}\cap I_{+},\mathbb{B}_{{+}}\cap
I_{+},_{\!}\xi )} \\
\!\!\! &=&\!\!\!1_{_{\!}{\Theta }_{\!+\!}(I_{+})}(\theta _{\!+\!})\!\left[
_{\!}\bar{\psi}_{_{\!}Z_{_{\!}+}}^{(\xi ,\theta _{_{\!}+})\!}\right]
^{\!I_{+}} \\
&\times &\!\!\!1_{\!\mathcal{F}_{_{\!}}(_{_{\!}}\mathbb{B}_{\!{+}\!})\!}(%
\mathbb{B}_{\!{+}\!}\cap \!I_{+})\text{ }r_{B,+}^{\mathbb{B}_{\!{+}\!}\cap
\!I_{+}}\left[ 1-r_{B,+}\right] ^{\mathbb{B}_{+}-(\mathbb{B}_{{+}}\cap
\!I_{+})} \\
&\times &\!\!\!\sum_{I}\!1_{\!\mathcal{F}\!(I)\!}(\mathbb{L}\!\cap
\!I_{+\!})\!\!\left[ _{\!}\bar{P}_{S\!}^{_{\!}(_{\!}\xi )\!}\right] ^{\!%
\mathbb{L}\cap \!I_{+}}\!\!\left[ _{\!}1\!-\!\bar{P}_{S\!}^{_{\!}(_{\!}\xi
)\!}\right] ^{\!I\!-(\mathbb{L}\cap \!I_{\!+\!})\!}\!\omega ^{(\!I,\xi )} \\
\!\!\! &=&\!\!\!1_{_{\!}{\Theta }_{\!+\!}(I_{+\!})}(_{\!}\theta _{\!+\!})\!%
\left[ _{\!}\bar{\psi}_{_{\!}Z_{_{\!}+}}^{(\xi ,\theta _{_{\!}+\!})\!}\right]
^{\!I_{+}}\!r_{B,+}^{\mathbb{B}_{{+}}\cap I_{+}}\left[ 1-r_{B,+\!}\right] ^{%
\mathbb{B}_{+}-I_{+\!}} \\
&\times &\!\!\!\sum_{I}\left[ _{\!}\bar{P}_{S\!}^{(\xi )\!}\right] ^{\!I\cap
I_{+\!}}\!\left[ _{\!}1\!-\!\bar{P}_{S\!}^{(\xi )\!}\right]
^{\!I\!-I_{+}\!}\omega ^{(I,\xi )}
\end{eqnarray*}%
where the last equality follows from $1_{\mathcal{F}(_{_{\!}}\mathbb{B}_{\!{+%
}\!})}(\mathbb{B}_{\!{+}}\cap I_{+})=1$, $\mathbb{B}_{+}-(\mathbb{B}_{\!{+}%
}\cap I_{+})=\mathbb{B}_{\!+\!}-I_{\!+}$, and $\mathbb{L}\cap I_{+}=I\cap
I_{+}$, $I\!-\!(\mathbb{L}\cap I_{+})=I\!-\!I_{\!+}$ for any $I_{+}$ such
that $1_{\mathcal{F}(I)}(\mathbb{L}\cap I_{+})=1$. Further, substituting the
above equation into (\ref{eq:GLMB_upd0}) and\ noting that $J$, $L_{\!+\!}$
are disjoint, the sum over the pair $J$, $L_{\!+}$ reduces to the sum over $%
I_{\!+}$. Hence, exchanging the order of the sums gives the desired result. $%
\square $

\medskip

\textbf{Proof of Lemma \ref{lemma}: }Note that $\gamma _{i}\!=\!\gamma
_{j}\!>\!0$ iff $\delta _{\gamma _{i\!}}[\gamma _{j}]1_{\{1:M\}}(\gamma
_{i})=1$. Hence, $\gamma $ is positive 1-1 iff for any distinct $i$, $j$, $%
\delta _{\gamma _{i\!}}[\gamma _{j}]1_{\{1:M\}}(\gamma _{i})=0$. Also, $%
\gamma $ is not positive \mbox{1-1} iff there exists distinct $i$, $j$ such
that $\delta _{\gamma _{i}\!}[\gamma _{j}]1_{\{1:M\}}(\gamma _{i})=1$.
Similarly, $\gamma _{\bar{n}\!}$ is positive 1-1 iff for any distinct $i$, $%
j\in \bar{n}$, $\delta _{\gamma _{i\!}}[\gamma _{j}]1_{\{1:M\}}(\gamma
_{i})=0$.

We will show that (a) if $\gamma $ is positive 1-1 then the right hand side
(RHS) of (\ref{eq:LemmaGibbs}) equates to 1, and \ (b) if $\gamma $ is not
positive 1-1, then the RHS of (\ref{eq:LemmaGibbs}) equates to 0.

To establish (a), assume that $\gamma $ is positive 1-1, then $\gamma _{\bar{%
n}\!}$ is also positive 1-1, i.e., $1_{{\Gamma }(\bar{n})}(\gamma _{\bar{n}%
\!})=1$, and for any $i\neq n$, $\delta _{\gamma _{n\!}}[\gamma
_{i}]1_{\{1:M\}}(\gamma _{n})=0$. Hence the RHS of (\ref{eq:LemmaGibbs})
equates to 1.

To establish (b), assume that $\gamma $ is not positive 1-1. If $\gamma _{%
\bar{n}\!}$ is also not positive 1-1, i.e., $1_{{\Gamma }(\bar{n})}(\gamma _{%
\bar{n}\!})=0$, then the RHS of (\ref{eq:LemmaGibbs}) trivially equates to
0. It remains to show that even if $\gamma _{\bar{n}\!}$ is positive 1-1,
the RHS of (\ref{eq:LemmaGibbs}) still equates to 0. Since $\gamma $ is not
positive 1-1, there exist distinct $i$, $j$ such that $\delta _{\gamma
_{i\!}}[\gamma _{j}]1_{\{1:M\}}(\gamma _{i})=1$. Further, either $i$ or $j$
has to equal $n$, because the positive 1-1 property of $\gamma _{\bar{n}\!}$
implies that if such (distinct) $i$, $j$, are in $\bar{n}$, then $\delta
_{\gamma _{i\!}}[\gamma _{j}]1_{\{1:M\}}(\gamma _{i})=0$ and we have a
contradiction. Hence, there exists $i\neq n$ such that $\delta _{\gamma
_{n\!}}[\gamma _{i}]1_{\{1:M\}}(\gamma _{n})=1$, and thus the RHS of (\ref%
{eq:LemmaGibbs}) equates to 0. $\square $

\medskip

\textbf{Proof of Proposition \ref{convergence}: }Convergence of finite state
Markov chains can be characterized in terms of irreducibility and
regularity. Following \cite{TaylorKarlin}, a Markov chain is \emph{%
irreducible} if it is possible to move from any state to any other state in
finite time, further, an irreducible finite state Markov chain is \emph{%
regular} if some finite power of its transition matrix has all positive
entries.

Consider the $n$th conditional $\pi _{n}(\gamma _{n}^{\prime }|\gamma
_{1:n-1}^{\prime },\gamma _{n+1:P})$, with $\gamma ^{\prime }$ positive 1-1.
Then for each $j\in \{1$:$n\!-\!1\}$, $1_{\{1:M\}}(\gamma _{j}^{\prime
})\delta _{\gamma _{n}^{\prime }}[\gamma _{j}^{\prime }]=0$, hence it
follows from (\ref{eq:marg_cond1}) that
\begin{eqnarray}
&&\!\!\!\!\!\!\!\!\!\!\!\!\!\!\!\!\!\!\!\!\!\!\!\!\!\!\!\!\!\!\!\!\!\!\pi
_{n}(\gamma _{n}^{\prime }|\gamma _{1:n-1}^{\prime },\gamma _{n+1:P})  \notag
\\
&=&\frac{\eta _{n}(\gamma _{n}^{\prime
})\prod\limits_{j=n+1}^{P}(1\!-\!1_{\{1:M\}}(\gamma _{j})\delta _{\gamma
_{n}^{\prime }}[\gamma _{j}])}{K_{n}(\gamma _{1:n-1}^{\prime },\gamma
_{n+1:P})}  \label{eq:convergence1}
\end{eqnarray}%
where $K_{n}(\gamma _{1:n-1}^{\prime },\gamma _{n+1:P})$ denotes the
normalizing constant in the $n$th sub-iteration of the Gibbs sampler.

Let $0_{n}$ denotes the $n$-dimensional zero vector. In addition (to being
positive 1-1), if $\gamma ^{\prime }=0_{P}$, then for each $j\in \{n\!+\!1$:$%
P\}$, $1_{\{1:M\}}(\gamma _{j})\delta _{0}[\gamma _{j}]=0$, because $\gamma
_{j}$ cannot be both positive and zero. Hence (\ref{eq:convergence1})
becomes $\eta _{n}(0)/K_{n}(0_{n-1}\mathbf{,}\gamma _{n+1:P})$ and since $%
\eta _{i}(j)$ is always positive (see definition (\ref{eq:eta})), we have
\begin{equation*}
\pi (0_{P}|\gamma )=\prod\limits_{n=1}^{P}\frac{\eta _{n}(0)}{K_{n}(0_{n-1}%
\mathbf{,}\gamma _{n+1:P})}>0.
\end{equation*}
On the other hand, if $\gamma =0_{P}$, then for each $j\in \{n\!+\!1$:$P\}$,
$1_{\{1:M\}}(\gamma _{j})=0$. Hence (\ref{eq:convergence1}) becomes $\eta
_{n}(\gamma _{n}^{\prime })/K_{n}(\gamma _{1:n-1}^{\prime },0_{P-(n+1)})$ and%
\begin{equation*}
\pi (\gamma ^{\prime }|0_{P})=\prod\limits_{n=1}^{P}\frac{\eta _{n}(\gamma
_{n}^{\prime })}{K_{n}(\gamma _{1:n-1}^{\prime },0_{P-(n+1)})}>0.
\end{equation*}%
Consequently, the probability of a 2-step transition from any $\gamma \in
\Gamma $ to any $\gamma ^{\prime }\in \Gamma $
\begin{equation*}
\pi ^{2}(\gamma ^{\prime }|\gamma )=\sum\limits_{\zeta \in \Gamma }\pi
(\gamma ^{\prime }|\zeta )\pi (\zeta |\gamma )>\pi (\gamma ^{\prime
}|0_{P})\pi (0_{P}|\gamma )>0.
\end{equation*}%
Hence, the chain is irreducible and also regular since the square of the
transition matrix has all positive elements.

Lemma 1 of \cite{Roberts_Gibbs94} asserts that for a finite state Gibbs
sampler, irreducibility (with respect to (\ref{eq:thetajoint_dis})) is a
sufficient condition for convergence to (\ref{eq:thetajoint_dis}). More
importantly, since the chain is regular, uniqueness of the stationary
distribution and the rate of convergence follows directly from \cite[Theorem
4.3.1]{Gallager14}, noting that $j=2$ is chosen since $\pi ^{2}$, the square
of the transition matrix, has all positive elements. $\square $

%

\providecommand{\url}[1]{#1} \csname url@samestyle\endcsname%
\providecommand{\newblock}{\relax} \providecommand{\bibinfo}[2]{#2} %
\providecommand{\BIBentrySTDinterwordspacing}{\spaceskip=0pt\relax} %
\providecommand{\BIBentryALTinterwordstretchfactor}{4}
\providecommand{\BIBentryALTinterwordspacing}{\spaceskip=\fontdimen2\font plus
\BIBentryALTinterwordstretchfactor\fontdimen3\font minus
  \fontdimen4\font\relax}
\providecommand{\BIBforeignlanguage}[2]{{\expandafter\ifx\csname l@#1\endcsname\relax
\typeout{** WARNING: IEEEtran.bst: No hyphenation pattern has been}\typeout{** loaded for the language `#1'. Using the pattern for}\typeout{** the default language instead.}\else
\language=\csname l@#1\endcsname
\fi
#2}} \providecommand{\BIBdecl}{\relax} \BIBdecl

\end{document}